\documentclass[aps,showpacs, nofootinbib,floatfix]{revtex4}
\usepackage{graphicx, graphics, color}

\usepackage[caption=false]{subfig}
\input amssym.tex
\input amssym.def

\newcommand{\be}{\begin{equation}}
\newcommand{\ee}{\end{equation}}
\newcommand{\ben}{\begin{eqnarray}}
\newcommand{\een}{\end{eqnarray}}

\newcommand{\la}{{\lambda}}

\newcommand{\cT}{{\cal T}}
\newcommand{\cO}{{\cal O}}

\newcommand{\cL}{{\cal L}}

\newcommand{\p}{\partial}
\newcommand{\na}{\nabla}

\newcommand{\tal}{{\tilde \alpha}}

\newcommand{\trho}{\tilde \rho}

\newcommand{\hM}{{\hat M}}
\newcommand{\hQ}{{\hat Q}}
\newcommand{\hht}{{\hat t}}
\newcommand{\hr}{{\hat r}}
\newcommand{\ha}{{\hat a}}
\newcommand{\hSigma}{{\hat \Sigma}}
\newcommand{\hDelta}{{\hat \Delta}}
\newcommand{\hmu}{{\hat \mu}}
\newcommand{\hrho}{{\hat \rho}}

\newcommand{\htheta}{\hat \theta}

\newcommand{\tB}{{\tilde B}}

\newcommand{\tT}{\tilde T}

\newcommand{\ep}{\epsilon}

\newcommand{\talpha}{\tilde \alpha}
\newcommand{\tbeta}{\tilde \beta}
\newcommand{\tla}{\tilde \lambda}

\newcommand{\ga}{\gamma}

\pacs{04.70.Bw, 98.80.Cq}

\begin{document}

\title{Abelian-Higgs hair on stationary axisymmetric black hole in Einstein-Maxwell-axion-dilaton
gravity}
%%%%%%%%%%%%%%%%%%%%%%%%%%%%%%%%%%%%%%%%%%%%%%%%%%%%%%%%%%%%%%

\author{{\L}ukasz Nakonieczny and Marek Rogatko}
\email{rogat@kft.umcs.lublin.pl, 
marek.rogatko@poczta.umcs.lublin.pl,
lnakonieczny@kft.umcs.lublin.pl}
\affiliation{Institute of Physics \protect \\
Maria Curie-Sklodowska University \protect \\
20-031 Lublin, pl.~Marii Curie-Sklodowskiej 1, Poland }

%\author{Marek Rogatko}
%\affiliation{Institute of Physics \protect \\
%Maria Curie-Sklodowska University \protect \\
%20-031 Lublin, pl.~Marii Curie-Sklodowskiej 1, Poland \protect \\
%rogat@kft.umcs.lublin.pl \protect \\
%marek.rogatko@poczta.umcs.lublin.pl}

%%%%%%%%%%%%%%%%%%%%%%%%%%%%%%%%%%%%%%%%%%%%%%%%%%%%%%%%%%%%%%%%%%%%
\date{\today}
%\pacs{04.30.Nk, 04.40.-b}

%%%%%%%%%%%%%%%%%%%%%%%%%%%%%%%%%%%%%%%%%%%%%%%%%%%%%%%%%%%%%%%%%%%%%%%%%%%%%%%%%%%%%%%%%%%%%%%%%%%
\begin{abstract}
We studied, both analytically and numerically, an Abelian Higgs vortex in the spacetime
of stationary axisymmetric black hole being the solution of the low-energy limit of the heterotic string theory,
acted as hair on the black hole background. Taking into account the gravitational backreaction
of the vortex we show that its influence is far more subtle than causing only a conical defect.
As a consequence of its existence, we found that the ergosphere was shifted as well as event horizon angular velocity
was affected by its presence. Numerical simulations reveal the strong dependence of the vortex
behaviour on the black hole charge and mass of the Higgs boson.
For large masses of the Higgs boson and large value of the charge, black hole is always pierced by an Abelian Higgs
vortex, while small values of the charge and Higgs boson mass lead to the expulsion
of Higgs field.

\end{abstract}
%%%%%%%%%%%%%%%%%%%%%%%%%%%%%%%%%%%%%%%%%%%%%%%%%%%%%%%%%%%%%%%%%%%%%%%%%%%%%%%%%%%%%%%%%%%%%%%%%%%%

\maketitle

%%%%%%%%%%%%%%%%%%%%%%%%%%%%%%%%%%%%%%%%%%%%%%%%%%%%%%%%%%%%%%%%%%%%%%%%%%%%%%%%%%%%%%%%%%%%%%%%%%%%%
\section{Introduction}
The {\it  no hair} conjecture of black holes, 
or its mathematical formulation, the uniqueness theorem of black holes states that the
static electrovacuum black hole spacetime is described by Reissner-Nordstr\"om solution whereas the circular one is 
diffeomorphic to Kerr-Newmann 
spacetime \cite{book}. The attempts of building a consistent quantum gravity theory 
and understanding the behaviour of matter field in a spacetime of higher-dimensional
black holes \cite{rog12} triggered the 
resurgence of works treating the mathematical aspects of higher-dimensional black objects.
In higher dimensional spacetime the uniqueness theorem for static black holes is well justified \cite{nd}, while
the stationary axisymmetric case is far more complicated (for the recent efforts of proving uniqueness theorem see, e.g.,
\cite{nrot}). Consequently, these researches comprise also the low-energy limit 
of the string theory, like dilaton gravity, Einstein-Maxwell-axion-dilaton gravity and 
supergravities theories \cite{sugra}. 
\par
In addition, black holes and 
their properties consist the key ingredients of the AdS/CFT attitude \cite{adscft} to superconductivity
acquire great attention. Questions about possible matter configurations in AdS spacetime 
arise naturally during aforementioned researches. In Ref.\cite{shi12}
it was shown that strictly stationary AdS spacetime could not allow for the existence of nontrivial
configurations of complex scalar fields or form fields. The generalization of the aforementioned problem,
i.e., strictly stationarity of spacetimes with complex scalar fields
in Einstein-Maxwell-axion-dilaton gravity with negative cosmological constant was given in \cite{bak13}.
\par
From the above point of view the other theories of gravity are also under intensive researches.
Namely, the strictly stationary static vacuum spacetimes in Einstein-Gauss-Bonnet theory were
discussed in \cite{shi13}, while
in Ref. \cite{shi13a} it was revealed that a static asymptotically flat black hole
solution is unique to be Schwarzschild spacetime in Chern-Simons-gravity.
Applying the conformal positive
energy theorem the uniqueness proof of static
asymptotically flat electrically charged black hole in dynamical Chern-Simons-gravity was performed \cite{rog13}. 
\par
The discovery of Bartnik and McKinnon \cite{ba88} of a nontrivial particle like 
structure in Einstein-Yang-Mills systems opens new realms of nontrivial solutions to 
Einstein-non-Abelian-gauge systems. It turned out that black holes can be colored \cite{ba88} - \cite{kun90}, support a 
long-range Yang-Mills hair. However, these solutions are unstable \cite{str90} - \cite{biz91}, but 
nevertheless they exist.
\par
%%%%%%%%%%%%%%%%%%%%%%%%%%%%%%%%%%%%%%%%%%%%%%%%%%%%%%%%%%%%%%%%%%%%%%%%%%%%%%%%%%%%%%%%%%%%%%
The other kind of problems is the extension of the aforementioned {\it no hair} conjecture to
the case when some field configurations have non-trivial topology. We may ask if topological defects \cite{vil}
can constitute hair on black hole spacetimes.
In Ref.\cite{ary86} the metric describing a Schwarzschild black hole threaded by a cosmic string
was provided. Then, the extension of the problem to the case of Abelian Higgs vortex on Euclidean Einstein
\cite{dow92} and Euclidean dilaton black hole systems \cite{mod98} were elaborated.
Further, numerical and analytical studies revealed that Abelian Higgs vortex could act as a long hair for
a Schwarzschild \cite{ach95} and Reissner-Nordstr\"om \cite{cha} black holes. The extremal Reissner-Nordstr\"om black hole
displays an analog of Meissner effect, but the flux expulsion does not occur in all
cases. The case of
charged dilaton black hole Abelian Higgs vortex system was treated in \cite{mod99}, where among all it was shown that all
extremal dilaton black holes always expelled vortex flux. On the other hand,
the problem of superconducting cosmic vortex and possible fermion condensations
around Euclidean Reissner-Nordstr\"om \cite{gre92} and dilaton black holes \cite{nak11} were elaborated,
while black string with superconducting cosmic string was studied in Ref.\cite{nak12}. 
\par
The case of stationary axisymmetric black hole vortex system turns out to be more complicated task.
The first attempts to attack this problem were presented in Refs.\cite{ghe02}, but the correct
treatment of Kerr black hole Abelian Higgs vortex was presented in \cite{gre13}.
\par
In our paper we shall provide some continuity with our previous works concerning
the dilaton static black hole topological defect systems \cite{mod98,mod99} as well as the works concerning 
the problem of the existence of
cosmic vortex in the spacetime of stationary axisymmetric black hole of Kerr type \cite{gre13}. Namely, we shall 
take into account charged stationary axisymmetric solution of the low-energy limit of the heterotic string theory,
Kerr-Sen black hole.
The uniqueness of Kerr-Sen black hole was proved in Ref.\cite{rog10} thus the next problem to consider will
be the question of possible hair on the black hole in question. In the light of the arguments 
presented in \cite{gre13} we shall 
look for the evidence that an Abelian Higgs vortex can act as a long hair for the Kerr-Sen black hole.
\par
The paper is organized as follows. In Sec.II we briefly review an Abelian Higgs vortex configuration
on stationary axisymmetric black hole solution in Einstein-Maxwell-axion-dilaton gravity. Then, we perform 
analytical considerations connected with the gravitational
backreaction problem in order to achieve the line element describing Kerr-Sen black hole pierced by
a vortex. We discuss its properties, especially the phenomenon of shifting the position of black hole
ergosphere due to the presence of the vortex. In Sec.III we present a numerical analysis
of the equations of motion for an Abelian Higgs vortex in the spacetime of the extremal and nonextremal
Kerr-Sen black hole. Sec.IV will be devoted to the conclusions of our investigations.

%%%%%%%%%%%%%%%%%%%%%%%%%%%%%%%%%%%%%%%%%%%%%%%%%%%%%%%%%%%%%%%%%%%%%%%%%%%%%%%%%%%%%%%%%%%%%%%%%%%%%%%%
%%%%%%%%%%%%%%%%%%%%%%%%%%%%%%%%%%%%%%%%%%%%%%%%%%%%%%%%%%%%%%%%%%%%%%%%%%%%%%%%%%%%%%%%%%%%%%%%%%%
%%%%%%%%%%%%%%%%%%%%%%%%%%%%%%%%%%%%%%%%%%%%%%%%%%%%%%%%%%%%%%%%%%%%%%%%%%%%%%%%%%%%%%%%%%%%%%%%%%%
\section{Einstein-Maxwell-axion-dilaton gravity}
In this section we shall study an Abelian Higgs vortex in the presence of Kerr-Sen black
hole being the stationary axisymmetric solution of the low-energy limit of heterotic string theory,
the so-called  Einstein-Maxwell-axion-dilaton gravity.
In our analysis we assume the complete separation of degrees of freedom for each
of the objects in question. The resulting action for the considered system will be the sum
of the action devoted to  Einstein-Maxwell-axion-dilaton gravity 
\be
S_1 = {1 \over 16 \pi G}~\int d^4 \sqrt{-g} \bigg[ R - 2~(\na \phi)^2 - {1 \over 2}~e^{4\phi} (\na a)^2
- e^{-2~\phi}F_{\mu \nu}F^{\mu \nu} - a~F_{\mu \nu}~{\ast F}^{\mu \nu} \bigg],
\ee
and the action for an Abelian Higgs model minimally coupled to gravity theory, which will be subject to 
spontaneous symmetry breaking. Its action implies the following:
\be
S_2 = \int d^4 \sqrt{-g}~\bigg[
- D_\mu \Phi^\dagger D^\mu \Phi - {1 \over 4}~\tB_{\mu \nu}~\tB^{\mu \nu} - {\tla \over 4}
\bigg( \Phi^\dagger \Phi - \eta^2 \bigg)^2 \bigg],
\ee
where we have denoted by $\Phi$ a complex scalar field. The covariant derivative is given by
$D_\alpha = \na_\alpha + i~e~B_\alpha$
and field strength tensor bounded with $B_\mu$ is of the form
$\tB_{\mu \nu} = 2~\na_{[\mu} B_{\nu]}$. Whereas the $U(1)$-gauge field strength tensor is given by
$F_{\alpha \beta} = 2~\na_{[\alpha}A_{\beta]}$. while the dual tensor to the $U(1)$-gauge field has the form 
$\ast F_{\alpha \beta} = 1/2~\ep_{\alpha \beta \ga \delta}~F^{\ga \delta}$.\\
As usual, we can define real fields 
$X,~P_\alpha,~\chi$ by the following relations:
\ben
\Phi(x_\mu) &=& \eta~X(x_\mu)~e^{i \chi(x_\mu)},\\
B_{\alpha}(x_\mu) &=& {1 \over e}~\bigg[ P_\alpha(x_\mu) - \na_\alpha \chi(x_\mu) \bigg].
\een
The above real fields represent the physical degrees of freedom of the broken symmetric phase.
Namely, $X$ is the scalar Higgs field, $P_\alpha$ the massive vector boson, while $\chi$ is a gauge
degree of freedom and it is not a local observable. However it can have a globally nontrivial
phase factor which indicate the presence of the vortex in question.
The equations of motion for $X$ and $P_\alpha$ fields are provided by
the relations
\ben
\na_\alpha \na^\alpha X &-& P_\beta~P^\beta~X - {\tla~\eta^2 \over 2}~\bigg( X^2 - 1 \bigg)~X = 0,\\
\na_\mu \tB^{\mu \nu} &-& 2 e^2 \eta^2~X^2~P^\nu = 0.
\een
On the other hand, the field equations for the considered system yield
\ben
G_{\mu \nu} &=& T_{\mu \nu} (F,~a,~\phi) + 8\pi G~\tT_{\mu \nu}(vortex),\\
\na_{\alpha}\na^{\alpha} \phi &-& {1 \over 2}~e^{4 \phi}~\na_\mu a~\na^\mu a + {1 \over 2}~e^{-2 \phi}~F_{\alpha \beta}
F^{\alpha \beta} = 0,\\
\na_\beta \bigg( e^{4 \phi}~\na^\beta a \bigg) &-& F_{\mu \nu}~{\ast F}^{\mu \nu} = 0,\\
\na_{\mu} \bigg( e^{-2~\phi}F_{\mu \nu}F^{\mu \nu} &+& a~{\ast F}^{\mu \nu} \bigg) = 0,
\een
Consequently, the energy momentum tensors for the adequate matter fields imply 
\ben
T_{\alpha \beta}(\phi) &=& 2~\na_\alpha \phi \na_\beta \phi - g_{\alpha \beta}( \na \phi )^2,\\
T_{\alpha \beta}(a) &=& {1 \over 2}~e^{4 \phi}~\na_\alpha a \na_\beta a - {1 \over 4} g_{\alpha \beta} (\na a)^2,\\
T_{\alpha \beta}(F) &=& 2~e^{-2 \phi} F_{\alpha \ga} F_{\beta}{}{}^{\ga} - {1 \over 2} e^{-2 \phi} g_{\alpha \beta}
F_{\mu \nu}F^{\mu \nu},
\een
while for the Abelian Higgs field one gets the following form
of the energy momentum tensor:
\be
\tT_{\mu \nu}(vortex) = 2~\eta^2 ~\na_\mu X \na_\nu X + 2~\eta^2 X^2~P_{\mu} P_{\nu} + {1 \over e^2} \tB_{\mu \alpha}
\tB_{\nu}{}{}^{\alpha} + g_{\mu \nu} \cL(\Phi,~B_\mu),
\ee
where $\cL(\Phi,~B_\mu)$ is the Lagrangian density connected with $\Phi$ and $B_\mu$ fields which yields
\be
\cL(\Phi,~B_\mu) = - \eta^2~\na_\mu X \na^\mu X - \eta^2~X^2~P_{\mu} P^{\mu} -
{1 \over 4 e^2} \tB_{\mu \nu} \tB^{\mu \nu} - {\tla~\eta^4 \over 4}~\bigg( X^2 - 1 \bigg)^2.
\ee
In order to proceed further we shall use
coordinates which contemplate the axial symmetry of the Kerr-Sen black hole Abelian Higgs vortex system.
Namely, we take into account Weyl form of the axisymmetric line element described by
\be
ds^2 = - e^{2 \la} ~dt^2 + \alpha^2~e^{-2 \la} ~\bigg( d \varphi + \beta~dt \bigg)^2 + e^{2(\nu - \la)} ~
\bigg( dx^2 + dy^2 \bigg),
\ee
where the functions under consideration are of $x,~y$ dependences.
Further, we define $x$ and $y$ coordinates by the relations
\be
x = \int {dr \over \sqrt{\Delta}}, \qquad y = \theta.
\ee
As far as the considered metric of stationary axisymmetric black hole 
solution of Einstein-Maxwell-axion-dilaton gravity is concerned it can be written 
in the form as \cite{gal95}
\ben \label{kerrsen}
ds^2 &=& - \bigg(
1 - {2~G~M~(r - r_m) \over \trho^2} \bigg)~dt^2 +
\trho^2~\bigg(
{dr^2 \over \Delta} + d \theta^2 \bigg)
+ {4~G~M~(r - r_m)~a~\sin^2 \theta \over \trho^2}~dt~d\varphi \\ \nonumber
&+&
\bigg[
r~(r - r_m) + a^2 + {2~G~M~(r - r_m)~a^2~\sin^2 \theta \over \trho^2} \bigg]
~\sin^2 \theta ~d \varphi^2,
\een
where we have denoted by $r_m= Q^2/G~M$ and the rest of the quantities are defined by
\be
\trho^2 = r~(r - r_m) + a^2~\cos \theta, \qquad
\Delta = (r-r_m)(r - 2 G M) + a^2.
\ee
One can remark that the above metric reduces to the static dilaton black hole solution \cite{gar92}
when
$a = 0$. On the other hand, when one puts $r_m = 0$, it contracts to the standard Kerr metric.
The other form of the rotating axion dilaton black hole was conceived in Ref.\cite{sen92},
but it turns out that by the adequate coordinate transformations and change of the parameters it can be
brought to the line element (\ref{kerrsen}). $M$ is the standard Schwarzschild mass while $a$ is the Kerr
rotation parameter related to the black hole angular momentum by $a = J/M$.
\par
Taking account of the corresponding Kerr-Sen line element, one gets the following correspondence
between Kerr-Sen metric and Weyl line element:
\be
\alpha_0 = \sqrt{\Delta}~\sin \theta, \qquad
e^{2 \la_0} = {\trho^2 ~\Delta \over \Sigma^2}, \qquad
\beta_0 = - {2~M~G ~a~(r - r_m) \over \Sigma^2}, \qquad
e^{2 \nu_0} = {\trho^4 ~\Delta \over \Sigma^2},
\ee
where $\Sigma^2 = [r~(r - r_m) + a^2]^2 - a^2~\Delta~\sin^2 \theta$.
\par
By virtue of the above one can rewrite the equations of motion for 
the Abelian Higgs vortex Kerr-Sen black hole system in the form as follows:
\be
R_{\alpha \beta} = \cT_{\alpha \beta} - {1 \over 2}~\cT~g_{\alpha \beta},
\ee
where the energy momentum tensor is provided by the expression
\be
\cT_{\alpha}{}{}^\beta = \ep~T_{\alpha}{}{}^\beta (vortex) + T_{\alpha}{}{}^\beta (F,~a,~\phi).
\ee
In the above equation, 
$T_{\alpha}{}{}^\beta (vortex)$ denotes the rescaled energy momentum tensor (see, e.g., \cite{dow92,mod99})
of the considered Abelian Higgs vortex,
while $\ep = 8 \pi G \eta^2$. Thus, the explicit forms of equations of motion in the underlying theory are
given by
\ben
%R_x{}{}^x + R_y{}{}^y = 
- {1 \over 2} \bigg[
2~\na^2 \alpha &-& \alpha^3~e^{- 4 \la}~(\na \beta)^2 + 4 \alpha \na^2 \nu + 4~e^\la ~\na(\alpha~\na e^{- \la})
\bigg] \\ \nonumber
&=& \sqrt{-g}~\bigg(( \cT_{x}{}{}^x - {1 \over 2}~\cT~\delta_{x}{}{}^x) +
( \cT_{y}{}{}^y - {1 \over 2}~\cT~\delta_{y}{}{}^y)\bigg),\\
% R_y{}{}^y = 
- {1 \over 2} \bigg[
- \alpha^3~e^{- 4 \la}~\beta^2_{,y} &+& 2 \alpha \na^2(\nu - \la) - 2~(\alpha_{y}~\nu_{y}
- \alpha_{x}~\nu_{x}) - 2~\na \alpha~\na \la + 4~\alpha~\la^2_{,y} + 2 ~\alpha_{,yy} \bigg]
\\ \nonumber
&=& \sqrt{-g}~( \cT_{y}{}{}^y - {1 \over 2}~\cT~\delta_{y}{}{}^y),\\
%R_\phi{}{}^t = 
- {1 \over 2} \bigg(
- 3~\na \alpha~\na \beta &+& 4~\alpha~\na \beta ~\na \la - \alpha~\na^2 \beta
\bigg)
= \sqrt{-g}~\cT_{\phi}{}{}^t,\\
%r_{\varphi}{}{}^{\varphi} =
{\alpha \over 2}~e^{- 2 \nu}~\bigg[
2~\na^2 \alpha &-& 2~\na \alpha~\na \la - 2~\alpha~\na^2 \la + \alpha^3 ~e^{- 4 \la}~(\na \beta)^2 \bigg]
= \cT_{\varphi \varphi} - {1 \over 2}~\cT~g_{\varphi \varphi},\\
%R_{t}{}{}^t + R_{\varphi}{}{}^\varphi =
- \na^2 \alpha &=&  \sqrt{-g}~\bigg(( \cT_{t}{}{}^t - {1 \over 2}~\cT~\delta_{t}{}{}^t) +
( \cT_{\varphi}{}{}^\varphi - {1 \over 2}~\cT~\delta_{\varphi}{}{}^\varphi)\bigg),
\een
where $\na = (\p_x,~\p_y)$. 
On the other hand, equations of motion for the matter fields imply
\ben
4~e^{4 \phi}~\alpha~e^{-2 \la + 2 \nu}~\bigg(
\na \phi~\na a~e^{2\la -2\nu} \bigg) &+& e^{4 \phi}~\na( \alpha~\na a) - 4~\p_yA_\varphi~\p_x A_t = 0,\\
\na (\alpha~\na \phi) &-& {1 \over 2}~e^{4 \phi}~(\na a)^2 + \alpha ~e^{-2 \phi -2 \la}
\bigg[ - \na A_t~(\na A_t - \beta~\na A_\varphi) \\ \nonumber
&-& \na A_\varphi~\bigg(- \na A_t - \na A_\varphi~\bigg({e^{4 \la} - \alpha^2~\beta^2 \over \alpha^2}\bigg)
\bigg) \bigg] = 0,\\
\na \bigg[
e^{-2 \phi -2\la}~\alpha~(\na A_t &-& \beta~A_\varphi)\bigg]
+ \p_x(a~A_{\varphi, y}) - \p_y(a~A_{\varphi, x}) = 0.
\een
In the 
considered case one should also take into account
$B_{\varphi}$ and $B_t$ components of the Abelian Higgs gauge field.
In what follows we shall apply an iterative procedure of solving equations of motion, expanding the equations
in terms of $\ep$. This is well justified because of the fact that, e.g., for the grand unified
theory string one has $\ep \leq 10^{-6}$.
Our starting background solution will be described by $X_0$ and $P_0$ \cite{gre13} 
and will constitute the Nielsen-Olesen vortex solution \cite{nil73}, provided by
\be
X \simeq X_{0}(R), \qquad P_{\varphi} \simeq P_{0}(R), \qquad
P_{t} \simeq \zeta~P_{\varphi}.
\ee
It can be easily found that $\zeta = - 2~M~G~a(r-r_m)/\rho^4$, where we put $\rho$ equal to
\be
\rho = r~(r-r_m) + a^2,
\ee
and $R = \rho~\sin \theta = \rho ~e^{- \la}$. Near the core of the cosmic string 
one gets that $\sin \theta \sim \cO(M^{-1})$, which results in
the relation $R^2_{,x} + R^2_{,y} \sim e^{2\nu_0 - 2\la_0}$.
This implies in turn that to the zeroth order in $\ep$, the components describing the energy momentum tensor
of Abelian Higgs vortex yield
%%%%%%%%
\ben
T_{(0)y}^{}{}^{y} &\simeq& {X'}_{0}^{2} - {X_{0}^2~P_{0}^2 \over R^2}
+ {\tbeta}~{{P'}_{0}^2 \over R^2} - {1 \over 4}~\bigg( X_0^2 - 1 \bigg)^2,\\
T_{(0)\varphi}^{}{}^{\varphi} &\simeq& - {X'}_{0}^2 + {X_{0}^2~P_{0}^2 \over R^2}
+ {\tbeta}~{{P'}_{0}^2 \over R^2} - {1\over 4}~\bigg( X_0^2 - 1 \bigg)^2,\\
T_{(0)t}{}{}^{t} &=& T_{x}{}{}^{x}
\simeq - {X'}_{0}^2 - {X_{0}^2~P_{0}^2 \over R^2}
- {\tbeta}~{{P'}_{0}^2 \over R^2} - {1\over 4}~\bigg( X_0^2 - 1 \bigg)^2,\\
T_{(0)xy} &\simeq& {\sqrt{\Delta} \over \rho}~(r-r_m)~R~\bigg(
2~{X'}_{0}^2 + {2~\tbeta}~{P'}_{0}^2 \bigg),\\
T_{(0)t}{}{}^{\varphi} &\simeq& 0,
\een
where 
%%%%%%%%%%%%%%%%%%%%%%%%%%%%%%%%
$\tbeta = {1/2 e^2}$ is the Bogomol'nyi parameter while $R = \sqrt{\tla}~\eta~\rho$.
A prime denotes differentiation with respect to $R$-coordinate.
\par
As in Refs.\cite{ach95} we expand the adequate quantities in series, i.e.,
$\alpha = \alpha_0 + \ep~\alpha_1$, etc. and solve the underlying equations of motion iteratively.
In the case under consideration except Abelian Higgs vortex fields one has to do with dilaton, axion and
Maxwell gauge field which appears at the level of $\ep^0$-order. It caused that to $\cO(\ep)$-order the
geometry of the problem in question is not only modified by the vortex fields but also one should take into 
account the backreaction of the rest of the matter fields appearing in EMAD-gravity.  
As we can see the energy momentum components are functions of $R$-coordinate and this leads us to the conclusion
that the modification of the Kerr-Sen stationary axisymmetric solution will also depend on this coordinate.
Therefore, we assume that the first order perturbed solutions of the equations of motion in the theory in question
will imply
\ben
\alpha_1 &=& \alpha_0~\tal_1(R), \qquad \la_1 = \la_1(R), \qquad \nu_1 = \nu_1(R), 
\qquad \beta_1 = \beta_1(R),\\ \nonumber
a_1 &=& a_1(R), \qquad \phi_1 = \phi_1(R), \qquad A^{(1)}_{\mu} = f(R)~A^{(0)}_{\mu},
\een
where $A^{(0)}_{\mu}$ denotes Maxwell field for Kerr-Sen solution.
Having in mind the form of the energy momentum tensor for cosmic vortex, near the string core, and the fact that
$(T_{t}{}{}^t - T_{\varphi}{}{}^\varphi - T)(F,~a,~\phi)$ is equal to zero, we conclude
that to the leading order one attains
\be
{d^2 \over d R^2} \tal_1 + {2 \over R}~{d \over d R} \tal_1 = 2~{X_{0}^2~P_{0}^2 \over R^2}
+ {1 \over 2}~\bigg( X_0^2 - 1 \bigg)^2.
\ee
It can be easily checked that $\tal_1(R)$ will be given by
\be
\tal_1(R) = \int_R^\infty ~{1 \over R^2}dR~\int_0^R~R'^{2}\bigg(
2~{X_{0}^2~P_{0}^2 \over R'^2}
+ {1 \over 2}~\bigg( X_0^2 - 1 \bigg)^2 \bigg)dR'.
\ee
On the other hand, the fact that $\beta$ is subdominant quantity and its derivatives  are also
subdominant, one reveals that $\ep^1$-order 
Einstein-Maxwell-axion-dilaton gravity equations of motion can be readily write down in the forms
\ben \label{rff}
R~\bigg(
R~\tal_1^{''} + 2~\tal_1^{'} &-& R~\la_1^{''} - \la_1^{'} \bigg) =
R^2~\bigg[
2~{X_{0}^2~P_{0}^2 \over R^2} + {\tbeta}~{(P_{0}^{'})^2 \over R^2}
+ {1 \over 4}~\bigg( X_0^2 - 1 \bigg)^2 \bigg] \\ \nonumber 
&+& R^2~\bigg[
{Q^2~f_0 \over \rho^4} + 4~f_0~{Q^2~a^2 \over \rho^4} \bigg],\\ \label{rxxyy}
%%%%%%%%%%%Rxx+Ryy%%%%%%%%%%%%%%%%%%%%%%%%%%%%%%%%%%%%%%%
\tal_1^{''} &+& 2~{\tal_1^{'} \over R} + 2~\nu_{1}^{''} - 2~\la_1^{''} - 2~{\la_1^{'} \over R} =
2~X_{0}^2 + {1 \over 2}~\bigg( X_0^2 - 1 \bigg)^2, \\ \label{ryy}
%%%%%%%%%%%%%%%Ryy%%%%%%%%%%%%%%%%%%%%%%%%%%%%%%%%%%%%%%%%%%%%%%%%%%%%%%%
- \tal_1^{''} &-& {\tal_1^{'} \over R} - \nu_{1}^{''} + \la_1^{''} + {\nu_1^{'} \over R}
+ {\la_1^{'} \over R} =
2~{X_{0}^2 + {\tbeta}~{(P_{0}^{'})^2 \over R^2} + {3 \over 4}~
\bigg( X_0^2 - 1 \bigg)^2 }\\ \nonumber
&+&
2~{f~Q^4 \over \rho^4} + 2~{Q^2~R~f^{'} \over \rho^5},\\
%%%%%%%%%%%%%%%%%%%%%Rxy%%%%%%%%%%%%%%%%%%%%%%%%%%%%%%%%%%%%%%%%%%%%%%%%%%%%%%%%%%%%%%%%%%%%%%%%%%%%%%%%%
- {\sqrt{\Delta} \over \rho}~\bigg(r &-& {r_m \over 2}\bigg)~
\bigg( R^2~\tal_1^{''} + 2~R~\tal_1^{'} - 4~\la_1^{'} \bigg) \simeq
{\sqrt{\Delta} \over \rho}~\bigg(r - {r_m \over 2}\bigg)~R~\bigg[
2~X_{0}^2 + {2~\tbeta}~(P_{0}^{'})^2 \bigg],\\
%%%%%%%%%%%%%%%%%%%%%%%%phi%%%%%%%%%%%%%%%%%%%%%%%%%%%%%%%%%%%%%%%%%%%%%%%%%%%%%%%%%%%%%%%%%%%%%%%%%%%%%%%%%%%%
\phi_1^{''} &+& {\phi_1^{'} \over R} - 2~{f~Q^2 \over \rho^4} + 2~{f~Q^2~a^2 \over \rho^4~R^2} \simeq 0,\\
%%%%%%%%%%%%%%%%%%%%%%%%%ax%%%%%%%%%%%%%%%%%%%%%%%%%%%%%%%%%%%%%%%%%%%%%%%%%%%%%%%%%%%%%%%%%%%%%%%%%%%%%%%%%
a_1^{''} &+& {a_1^{'} \over R} - 4~{\phi_1^{'} \over \rho^4} - {r_m~a~\tal_1 \over \rho^4} \simeq 0,\\ \label{ff}
%%%%%%%%%%%%%%%%%%%%%%%%%%%%%A%%%%%%%%%%%%%%%%%%%%%%%%%%%%%%%%%%%%%%%%%%%%%%%%%%%%%%%%%%%%%%%%%%%%%%%%%%%%
f^{''} &+& {f^{'} \over R} \simeq -2~{\rho \over R~r}~(\phi_1^{'} + \la_1^{'}) + 2~{\tal_1 \over R^2~r}.
\een
As in Refs.\cite{mod98}-\cite{mod99} we
shall work in the so-called {\it thin string limit}. It means that one assumes that
the mass of the black hole in question is subject to the inequality
$M \gg 1$. Thus we shall neglect terms of the order $\cO(1/M^{n\ge 2})$. 
First, let us consider equation (\ref{rff}) and relation for $\talpha_1$. They give us the following
expression for $\la_1$:
\be
\la_1 \simeq \int_R^\infty {dR \over R}~\int_0^R dR'~{R'}~\bigg[
{1\over 4}~\bigg( X_0^2 - 1 \bigg)^2 - {\tbeta}~{(P_{0}^{'})^2 \over
{R'}^2} \bigg].
\ee
Hence, from equation (\ref{ff}), dropping terms of order 
$\cO(M^{-2})$,
one concludes that 
if we put the first integration constant to zero, then finally $f = f_0$, where $f_0$ is a constant value.
Turning 
our attention to the relations (\ref{rxxyy}) and (\ref{ryy}), we observe that $\nu_1 = 2~\la_1$. Then,
from equation describing the dilaton field we find that
\be
\phi_1 = \int_R^\infty dR~{2 \over R}~\int_0^R dR'~{f_0~Q^4 \over \rho^4}~\bigg(
R' - {a^2 \over R'} \bigg).
\ee
Referring our studies to the relations describing
axion field, it entails that the
following is satisfied:
\be
a_1 = \int_R^\infty dR~{1 \over R}~\int_0^R dR'~{R' \over \rho^4}~\bigg(
4~\phi_1^{'} + r_m~a~\tal_1 \bigg).
\ee
The essential point, however, is that the first order correction of $\beta$ cannot be established from an
asymptotical analysis, due to the fact that $\beta$-quantity is a subdominant function in the
problem in question. Just, taking the divergent part of the Ricci curvature tensor $R_{xy}$ one 
has that to $\ep^1$-oder we arrive at the relation
\be
e^{-4\la_0}~\alpha_0^2~\beta_{0,x}~\beta_{1,y} \simeq
{2~M~a~R^2 \over \rho^2~\sqrt{\Delta}} ~\bigg(
-1 + {4~r~(r-r_m) \over \rho^2} + \cO \bigg( {1 \over \rho^4} \bigg)\bigg)~\beta_{1,y}.
\ee
On this account it is customary to
examine the right-hand side of equation in question,
i.e., studying the adequate components of the energy momentum tensor, both for Abelian Higgs vortex and matter fields,
one infers that we cannot find $\delta \beta$ which has the
mandatory functional dependence on the coordinates.
It is remarkable fact that, the required form of $\ep^1$-order $\beta$ correction required for a pure
$\varphi$ deficit angle leads to the divergence of $xy$-component of the Ricci curvature tensor
at the event horizon of the considered black hole. In view of these arguments, we deduce that 
$\delta \beta = 0$.
\par
After rescaling coordinates
$\hht = e^{\ep~\nu_1/2}t, ~\hr = e^{\ep~\nu_1/2}r, ~\ha =  e^{\ep~\nu_1/2}a,~\hM =  e^{\ep~\nu_1/2}M,~\hQ=
 e^{\ep~\nu_1/2}Q $, one achieves to the line element describing {\it thin string} in Kerr-Sen black hole spacetime.
The resultant metric is provided by
\ben
ds^2 &=& - \bigg( {\hDelta - \ha^2~\sin^2 \theta \over \hrho^2} \bigg)~d \hht^2 
- {8~G^2~\hM^2~\ha^2~(\hr - \hr_m)^2 \over \hSigma^4}~\ep~\hmu~\bigg[
\hr~(\hr - \hr_m) + \ha^2 \bigg]~\sin^2 \theta~d \hht^2 \\ \nonumber
&+& {\hSigma^2 \over \hDelta}~d \hr^2 + \hrho^2~d \htheta^2 + {\hSigma^2~\sin^2 \theta \over \hrho^2}
\bigg( 1 - 2 ~\ep~\hmu \bigg)~d \varphi^2 
- {4~G~\hM~\ha~(\hr - \hr_m) \over \hrho^2}~\sin^2 \theta~\bigg( 1 - 2 ~\ep~\hmu \bigg)~d \hht~ d\varphi,
\een
where we have denoted by $\hmu$ mass per unit length of the cosmic string \cite{dow92}, while 
the other quantities are defined as
follows:
\ben
\hDelta &=& (\hr - \hr_m)~(\hr - 2~G~\hM) + \ha^2,\\
\hrho^2 &=& \hr~(\hr - \hr_m) +\ha^2~\cos^2 \theta,\\
\hSigma^2 &=& \bigg[ \hr~(\hr - \hr_m) +\ha^2 \bigg]^2 - \ha^2~\hDelta^2~\sin^2 \theta, \qquad 
\hr_{m} = {\hQ^2 \over G~\hM}.
\een
One can remark that the Abelian Higgs vortex on Kerr-Sen rotating black hole causes
not only an angular deficit angle which is felt by $\varphi$-coordinate but also
the deficit which influences $t$-coordinate. This fact leads us to a quite new physical phenomenon. Namely,
because of the fact that the deficit angle appears in $g_{tt}$ and the condition on the radius of the ergosphere
is $g_{tt}(r_{erg}) = 0$, then the position of the ergosphere is shifted
(we shall pay more attention to this problem in Sec.III). The same occurrence takes place
in Kerr Abelian Higgs vortex system \cite{gre13}. 
One should remark that recently the problem of ergoregions attracted much attention, see, e.g.,
\cite{gib13} and references therein.
\\
The angular velocity of the observer in Kerr-Sen black hole Abelian Higgs vortex system belongs to the interval
$\Omega_{min} <\Omega<\Omega_{max}$, where $\Omega_{{max},{min}} = \omega 
\pm \sqrt{\omega^2 - g_{tt}/g_{\varphi \varphi}}$, with $\omega = - g_{\varphi t}/g_{\varphi \varphi}$.
Thus, it is affected by the presence of the Abelian Higgs vortex. At the event horizon of the black hole
threaded by a vortex, where $\omega^2 = g_{tt}/g_{\varphi \varphi}$, 
$\Omega_{min}$ coincides with $\Omega_{max}$. The limiting angular velocity is of the form
\be
\Omega_h = \omega^2(r_{h},~\theta) = \omega^2_{Kerr-Sen}~\bigg[
1 + \ep~\hmu~\bigg(2 - 8~G~\hM~(\hr - \hr_m) \bigg) \bigg],
\ee
where we denoted
\be
\omega^2_{Kerr-Sen} = {\ha^2 \over [ \hr~(\hr - \hr_m) + \ha^2]}.
\ee
The limiting angular velocity $\Omega_h$, sometimes called the angular velocity
of the event horizon, is also modified by the presence of the Abelian Higgs vortex.\\ 
%%%%%%%%%%%%%%%%%%%%%%%%%
It is worth mentioning that, when we consider the area of event horizon of the the Kerr-Sen 
black hole penetrated by an Abelian Higgs vortex
it satisfies up to $\ep^1$-order
\be
A_{Kerr-Sen-vortex} = \int \int {\sqrt{g_{\varphi \varphi}~g_{\theta \theta}}~d \varphi~d \theta}
\simeq
A_{Kerr-Sen}~\sqrt{1 - 2 ~\ep~\hmu},
\ee
which means that it is affected by the vortex presence.

%%%%%%%%%%%%%%%%%%%%%%%%%

%%%%%%%%%%%%%%%%%%%%%%%%%%%%%%%%%%%%%%%%%%%%%%%%%%%%%%%%%%%%%%%%%%%%%%%%%%%%%%%%%%%%%%%%%%%%%%%%%%%%%%%%%%%
%%%%%%%%%%%%%%%%%%%%%%%%%%%%%%%%%%%%%%%%%%%%%%%%%%%%%%%%%%%%%%%%%%%%%%%%%%%%%%%%%%%%%%%%%%%%%%%%%%%%%%%%%%%
\section{Numerical results}
In this section one performs a number of numerical calculations to study the behaviour
of the Abelian Higgs vortex in Kerr-Sen spacetime. We commence our studies
with introducing dimensionless quantities useful in numerical solutions of the underlying equations.
Namely, one establishes the following rescaling:
\ben
(\tilde{Q}, \tilde{a}, \tilde{r}, 1) &=& \frac{1}{GM}( Q, a, r, GM), \\
(\tilde{m_{X}}, \tilde{m_{V}} ) &=&  GM( m_{X}, m_{V}), \\
(\tilde{P^{\phi}}, \tilde{P^{t}}) &=& (G^2M^2 P^{\phi}, GM P^{t}),
\een
where 
we have denoted the Higgs boson mass by
$m_{X} = \sqrt{\tla}~ \eta$ and 
the mass of the vector field in the broken phase by
$m_{V} = \sqrt{2} e \eta$.
In what follows, for the brevity of our notation, one drops the tilde from the rescaled quantities.
\par
It may be recalled that, in analogous to Kerr-Newmann solution in Einstein-Maxwell gravity, Kerr-Sen solution
is characterized by it mass, electric charge $Q$ and rotation parameter $a$ related to the black hole 
angular momentum and mass.
The maximal value of the rotation parameter can be inferred from the condition for the extremal black hole. Namely,
one has that the outer and inner horizons coalesce $r_h = r_+ = r_-$, where $r_{\pm} = 
G~M+ {r_m \over 2} \pm \sqrt{(G~M - {r_m \over 2})^2 - a^2}$. Hence, in our rescaling $a_{max} = 1 - \frac{1}{2}Q^2$.
Consequently, the maximal value of the black hole charge is fixed by the relation
$r_h = r_{-} = r_{m} = Q^2$. It leads to the conclusion that $Q_{max} = \sqrt{2}$.

%%%%%%%%%%%%%%%%%%%%%%%%%%%%%%%%%%%%%%%%%%%%%%%%%%%%%%%%%%%%%%%%%%%%%%%%%%%%%%%%%%%%%%%%%%%%%%%%%
\subsection{Location of the ergosphere - numerical results}
\begin{figure}[H]
\subfloat[]{\label{fig1a}\includegraphics[scale=1.5,angle= -90]{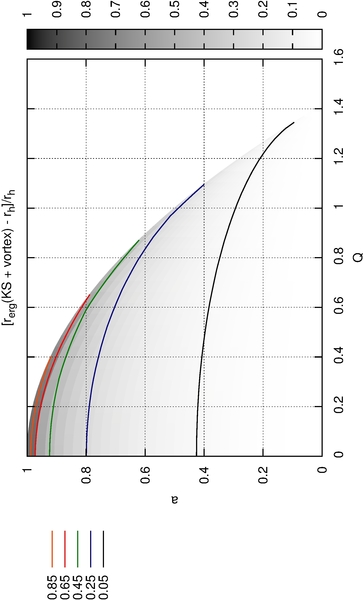}} \quad
\subfloat[]{\label{fig1b}\includegraphics[scale=1.5,angle= -90]{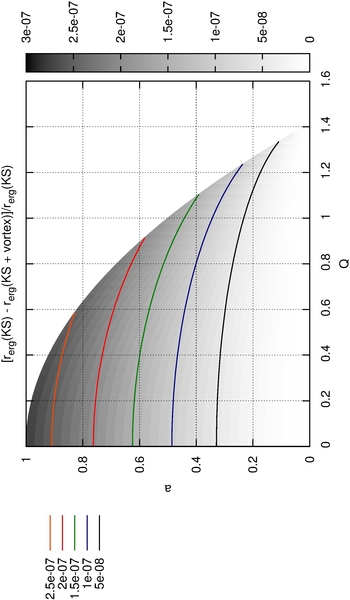}}
\caption{(color online) The difference between locations of (a) the ergosphere and the black hole event horizon
for the Kerr-Sen black hole vortex system and 
(b) the ergosphere for the pure Kerr-Sen black hole and for the Kerr-Sen black hole vortex system. 
We set $\theta = \frac{\pi}{2}$ and $\epsilon = 10^{-6}$.}
\end{figure}

To begin with we pay more attention to the problem of the ergosphere shifting due to the presence
of Abelian-Higgs vortex. On this account, in
Fig.1a we 
plotted the distance between the event horizon and the ergosphere for
Kerr-Sen Abelian Higgs vortex system, as a function of black hole charge $Q$ and angular momentum parameter $a$. 
We fix $\ep = 10^{-6}$ and 
perform this figure and the subsequent ones in the equatorial plane for which $\theta = \pi/2$. It may be 
seen that the increase of the value of black hole charge causes the diminishing
of the distance in question and in the case of $Q = Q_{max} = \sqrt{2}$ the ergosphere collapses onto
the Kerr-Sen event horizon.\\
The difference between the location of the ergosphere for pure Kerr-Sen black hole and Kerr-Sen black hole
pierced by an Abelian Higgs vortex is presented in Fig.\ref{fig1b}. It was done as a function of black hole charge.
One can conclude that 
the sign of the difference indicates that 
after taking into account the presence of the vortex
the ergosphere is shifted towards the black hole event horizon. The magnitude of this shift is 
of order equal to $0.1 \epsilon$. 
\par
\begin{figure}[H]
\subfloat[]{\label{fig2a}\includegraphics[scale=1.5,angle = - 90]{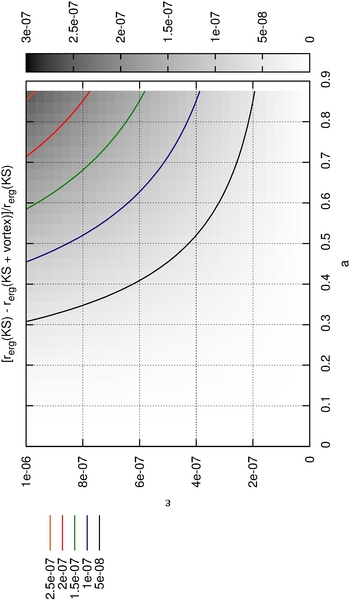}} \quad
\subfloat[]{\label{fig2b}\includegraphics[scale=1.5,angle = - 90]{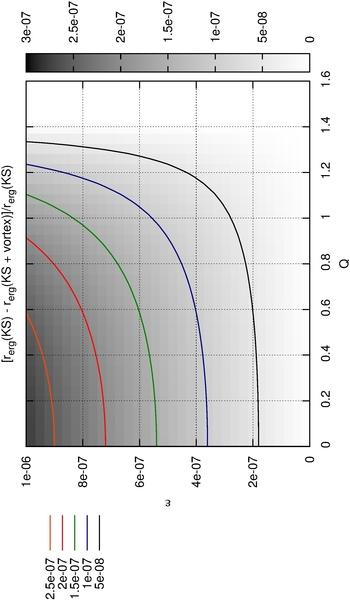}}
\caption{(color online) The difference between locations of (a) the ergosphere for the pure Kerr-Sen black hole
and for the Kerr-Sen black hole vortex system as a function of angular momentum parameter $a$ for fixed black hole charge 
$Q = 0.5Q_{max}$, 
(b) the ergosphere for the pure Kerr-Sen black hole 
and for the Kerr-Sen black hole vortex system as a function of $Q$ 
for $a = a_{max}$ (extremal black hole).
One considers the equatorial plane for which $\theta = \frac{\pi}{2}$.}
\end{figure}
\begin{figure}[H]
\subfloat[]{\label{fig3a}\includegraphics[scale=1.5,angle = - 90]{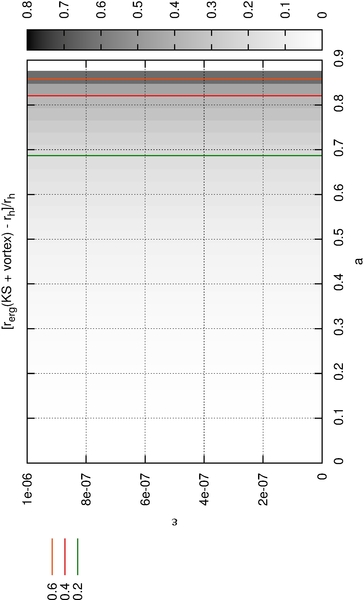} } \quad
\subfloat[]{\label{fig3b}\includegraphics[scale=1.5,angle = - 90]{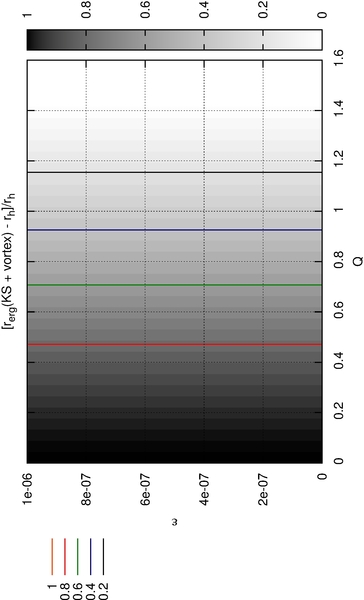}}
\caption{(color online) The difference between locations of (a) the ergosphere and the black hole horizon
for the Kerr-Sen vortex system as a function of $a$ for $Q = 0.5 Q_{max}$, 
(b) the ergosphere and the black hole horizon
for the Kerr-Sen vortex system as a function of black hole charge $Q$ for $a = a_{max}$ (extremal black hole).
The equatorial plane $\theta = \frac{\pi}{2}$ was taken into account.}
\end{figure}

In Fig.\ref{fig2a} we depicted the difference between the location of the ergosphere for Kerr-Sen black hole and the 
ergosphere of Kerr-Sen Abelian Higgs vortex system, as a function of $\ep$ and 
angular momentum parameter $a$, for fixed value of the black 
hole charge equal to $Q = 0.5~Q_{max}$. It can be observed that 
for the fixed $\epsilon$ and the black hole charge $Q$, the shift increases in its magnitude as the considered black 
hole approaches extremality. In Fig.\ref{fig2b} 
the difference between the location of the ergosphere for Kerr-Sen black hole and the 
ergosphere of Kerr-Sen Abelian Higgs vortex system, as a function of $\ep$ and $Q$ is plotted, for the maximal value 
of the angular momentum parameter.
%%%%%%%%%%%%%%%%%%%%%% 
From this figure we infer that the shift decreases as the black hole charge approaches the maximal one.
%%%%%%%%%%%%%%%%%%%%%%
The angular momentum parameter $a$ is fixed but the relation between the maximal allowed $a$ and $Q$ implies that
the bigger $Q$ is the smaller $a$ one obtains. 
Summing it all up, one can draw a conclusion that the 
increase of the value of black hole charge $Q$ (decrease of $a$) influences the decrease 
in the magnitude of the shift in the ergosphere position.
On the other hand, Fig.\ref{fig3a} and Fig.\ref{fig3b} indicate that for fixed values of the black hole charge $Q$ and 
angular momentum parameter $a$, the 
distance between the black hole event horizon and the ergosphere is insensitive to the modification of the value of $\ep$ 
(in the considered range of this parameter).

%%%%%%%%%%%%%%%%%%%%%%%%%%%%%%%%%%%%%%%%%%%%%%%%%%%%%%%%%%%%%%%%%%%%%%%%%%%%%%%%%%%%%%%%%%%%%%%%%%%%%%%%%%%%%%%%%%%
\subsection{Numerical solution of the equations of motion for Kerr-Sen Abelian Higgs vortex system}
In this subsection we shall analyze numerically equations of motion for an Abelian Higgs vortex
on the background of Kerr-Sen black hole. Namely, one considers the following relations:
\ben
\label{system_X}
\na_\mu \na^\mu X  &-& P^2 X - \frac{m_{X}^2}{2} X (X^2 - 1) = 0, \\
\na_\mu \na^\mu P^{\alpha} &-& m_{V}^2~ P^{\alpha} X^2 - 
R^{\alpha}_{~~ \beta}P^{\beta} = 0, \qquad \alpha = \phi,~ t.
\label{system_P}
\een
The relevant components of the Ricci curvature tensor are as follows: $R^{t}_{~~ t}$, 
$R^{t}_{~~ \varphi}$, $R^{\varphi}_{~~ \phi}$ and $R^{\varphi}_{~~ t}$, while their explicit forms imply
\ben
R^{t}_{~ t} &=& - 
\frac{G~M~[ r~(r - r_{m}) + a^2~(1 + \sin^2 \theta)]~[ (r - r_{m})^2 + a^2~ \cos^2 \theta]~r_{m}}{ \trho^8}, \\
%%%%%%%%%%%%
R^{t}_{~ \varphi} &=& 
\frac{ 2~G~M~a~[ (r - r_{m})^2 + a^2 ~\cos^2 \theta]~[r~(r-r_{m}) + a^2 ]~r_{m}~ \sin^2 \theta}{\trho^8}, \\
%%%%%%%%
R^{\varphi}_{~~ t} &=&
 - \frac{ 2~G~M~a~[ (r - r_{m})^2 + a^2~ \cos^2 \theta]~r_{m}}{\trho^8}, \\
%%%%%%%%%%%%
R^{\varphi}_{~~ \varphi} &=&
\frac{ G~M~ [ (r - r_{m})^2 + a^2~ \cos^2 \theta]~ [r ~(r - r_{m}) + a^2~(1 + \sin^2 \theta)]~r_{m}}{\trho^8}.
\een
The equations of motion (\ref{system_X})-(\ref{system_P})
are elliptic in the Kerr-Sen background, while on the black hole event horizon
they are parabolic. In order to solve them one uses the standard iterative technique. In the case under
consideration, we use the Newton method \cite{Press}. Moreover, a rectangular
grid in the $(r,\theta)$-plane will be exploited, where the range of the coordinates was chosen as
$[r_{h},~ 10~r_{h}] \times [0,~\pi]$. At large radii, one requires to obtain Nielsen-Olesen
solution, then the asymptotic values of the functions $X$ and $P^{\varphi}$ yield
\ben
(X,~ P^{\varphi}) &=& (0,~1) \qquad \theta = 0,~ \pi, \\
(X,~ P^{\varphi}) &=& (1,~0) \qquad r = 10~ r_{h}.
\een
On the other hand, we have a freedom in choosing the 
adequate boundary condition for $P^{t}$ component of the gauge field. 
A close inspection of the equation of motion satisfied by $P^{t}$ reveals that it contains the term
$\frac{1}{\sin(\theta)} \partial_{\theta} P^{t}$, which 
becomes singular at $\theta = 0,~ \pi$
unless $(\partial_{\theta} P^{t})_{| \theta = 0, \pi} = 0$.
From the physical point of view $P^{t}$ should vanish at infinity. In order to fulfill these requirements
one chooses that $P^{t}(\theta = 0,~ \pi) = 0$, as our boundary conditions.\\
In terms of the above we arrive at the following set of the boundary conditions for the fields 
composing the Abelian Higgs vortex:
\ben
(X,~ P^{\varphi},~ P^{t}) = (0,~1,~0) \qquad \theta = 0,~ \pi, \\
(X,~ P^{\varphi},~ P^{t}) = (1,~0,~0) \qquad r = 10~ r_{h}.
\een  
As in the previously considered cases of the Abelian Higgs vortex on various black hole backgrounds 
\cite{ach95}-\cite{mod99}, we  initialize our numerical simulations by choosing initial profiles 
for $X$ and $P^{\varphi}$. We commence with the boundary conditions on Kerr-Sen
black hole event horizon given by $X(r_{h}) \sim \sin^2 \theta$ and $P^{\varphi} \sim \cos^2 \theta$. 
Initially, we set everywhere $P^{t} = 0$. Next, few iterations with the Gauss-Seidel method \cite{Press}
were performed in order to achieve the appropriate starting profile for the Newton method. 
\par
%%%%%%%%%%%%%%%%%%%%%%%%%
It is worth mentioning that we may further reduce the number of the parameters describing our problem 
by setting the Bogomol'nyi parameter $\tbeta = \frac{m_{X}^2}{m_{V}^2}$  as equal to unity.  
$\tbeta = 1$ 
%%%%%%%%%%%%%%%%%%%%%%%%
corresponds to the situation when the magnetic and the Higgs flux tubes are of the same sizes,
sometimes one calls it that the vortex is supersymmetrizable. 
This fact allows us to examine the gauge field and  the Higgs boson mass
as fixed. In our rescaling it implies that modification of $m_{X}$ value is equivalent to the changes
of the black hole mass. Hence, one performs the numerical simulations for various values of black hole charges
of $Q$, boson Higgs masses $m_{X}$ and angular momentum parameters $a$.
\par
\begin{figure}[H]
\subfloat[$X(Q = 0.1 Q_{max}, a = 0.5 a_{max})$]{\includegraphics[scale=1.2]{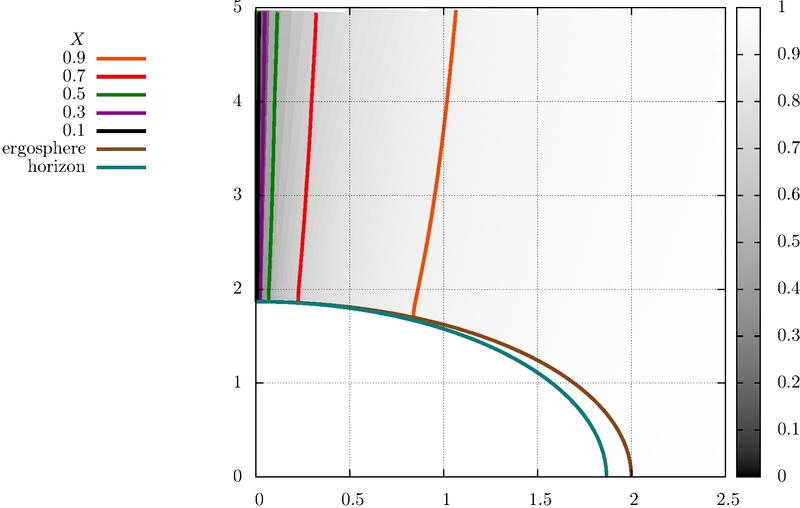}} \qquad
\subfloat[$P^{\varphi}(Q = 0.1 Q_{max}, a = 0.5 a_{max})$]{\includegraphics[scale=1.2]{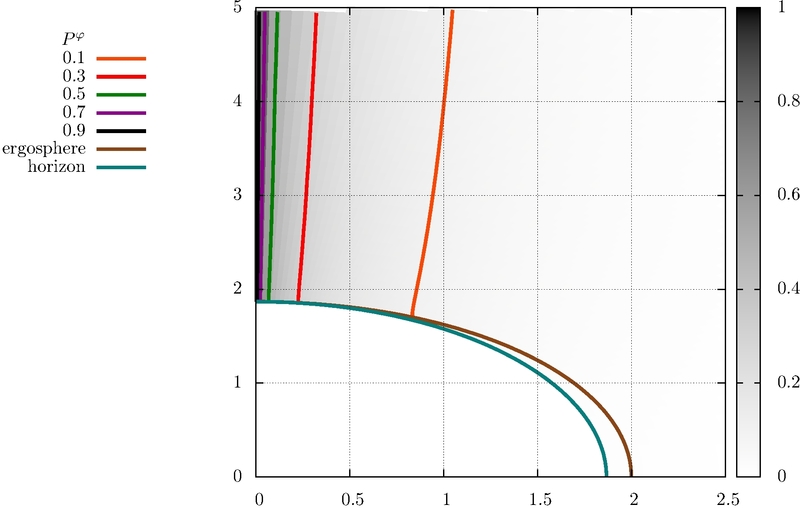}}
\\
\subfloat[$P^{t}(Q = 0.1 Q_{max}, a = 0.5 a_{max})$]{\includegraphics[scale=1.2]{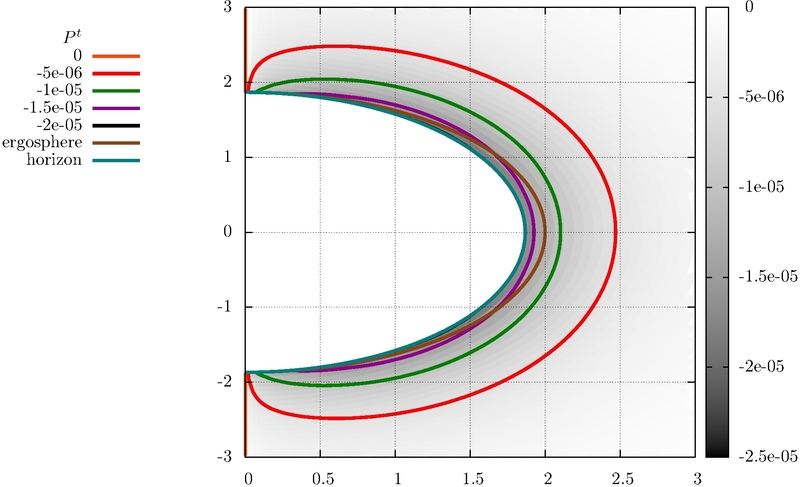}}
\caption{(color online) Behavior of the Higgs and gauge fields in the background of the nonextremal 
Kerr-Sen black hole.
One puts the boson Higgs mass equal to  $m_{X} = 0.9$.}
\label{fig4}
\end{figure}
\begin{figure}[H]
\subfloat[$X(Q = 0.1 Q_{max}, a = 0.9 a_{max})$]{\includegraphics[scale=1.2]{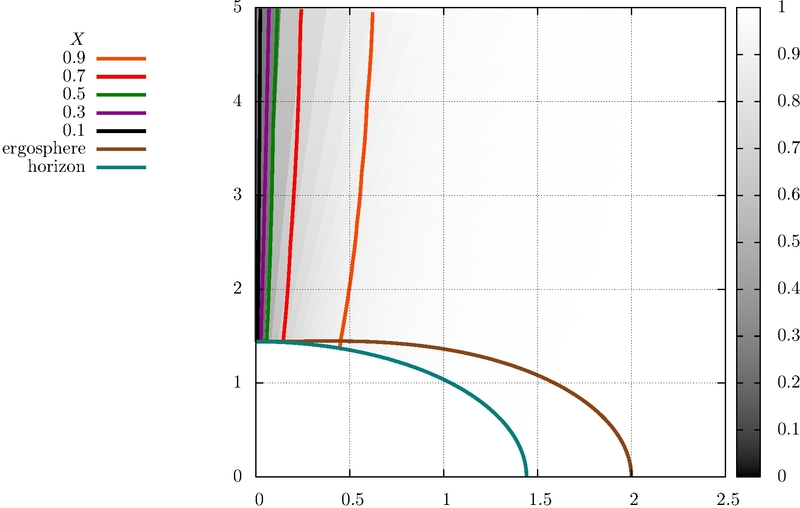}} \qquad
\subfloat[$P^{\varphi}(Q = 0.1 Q_{max}, a = 0.9 a_{max})$]{\includegraphics[scale=1.2]{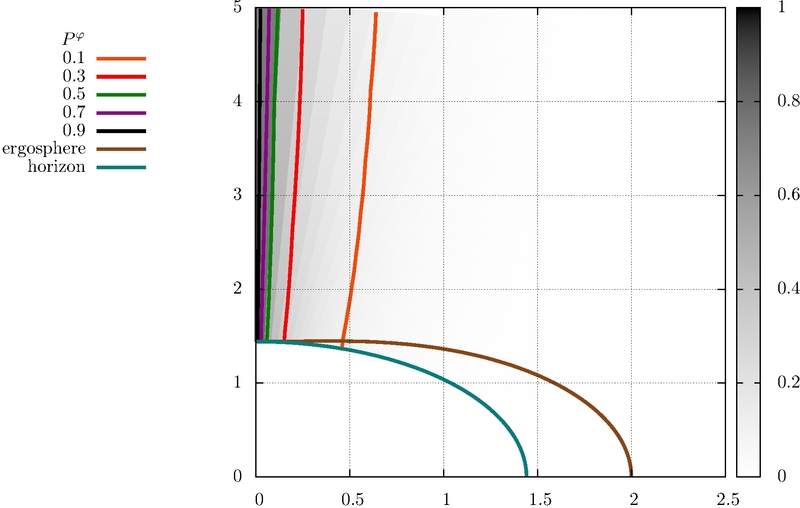}}
\\
\subfloat[$P^{t}(Q = 0.1 Q_{max}, a = 0.9 a_{max})$]{\includegraphics[scale=1.2]{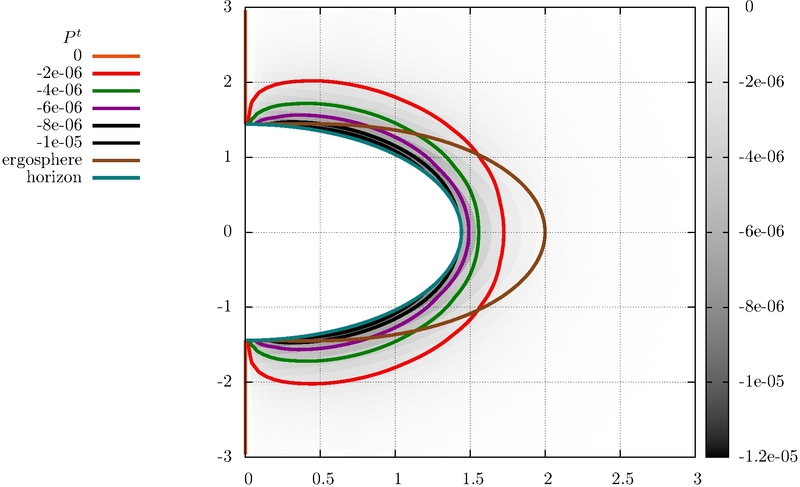}}
\caption{(color online) Behavior of the Higgs and gauge fields 
in the spacetime of the nonextremal Kerr-Sen black hole.
We fix the boson Higgs mass as equal to $m_{X} = 4$.}
\label{fig5}
\end{figure}

At the beginning we shall elaborate the case of the nonextremal Kerr-Sen black hole Abelian Higgs vortex system.
Namely, in Fig.\ref{fig4} we set $Q = 0.1~Q_{max}$ and $a = 0.5~a_{max}$, 
the Higgs boson mass we choose as $0.9$. On the other hand,
in Fig.\ref{fig5} one puts $Q = 0.1~Q_{max}$, $a = 0.9~a_{max}$ and $m_{X} = 4$, respectively.
These figures enable us to conclude that in the case of the nonextremal Kerr-Sen black hole
the Abelian Higgs vortex always pierces the black hole in question. Moreover, one can notice that
the increase of the Higgs boson mass causes the decrease of the width of the considered vortex.
Consequently, the larger angular parameter we consider the closer to the Kerr-Sen black hole
event horizon $P^{t}$ field is concentrated.
\par
\begin{figure}[H]
\subfloat[$X$]{\includegraphics[scale=1.2]{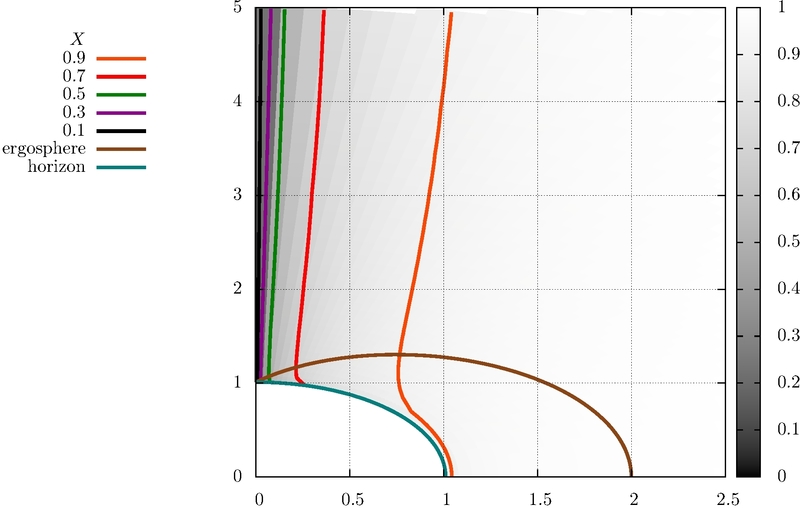}} \qquad
\subfloat[$P^{\varphi}$]{\includegraphics[scale=1.2]{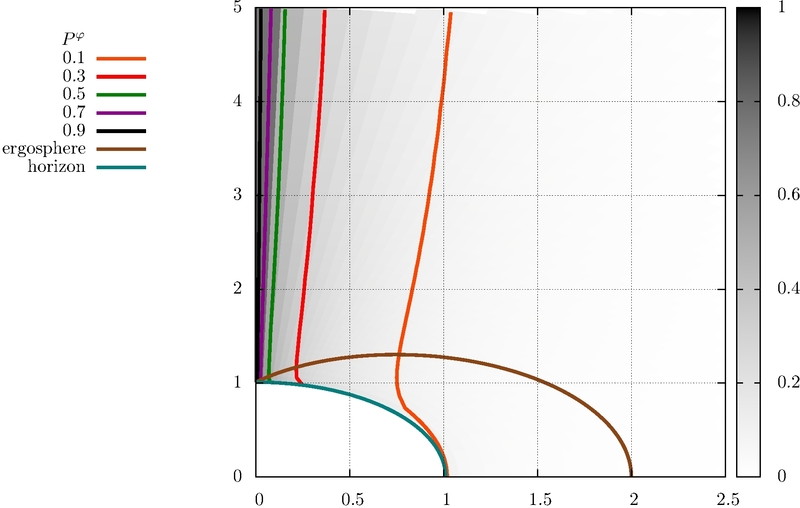}}
\\
\subfloat[$P^{t}$]{\includegraphics[scale=1.2]{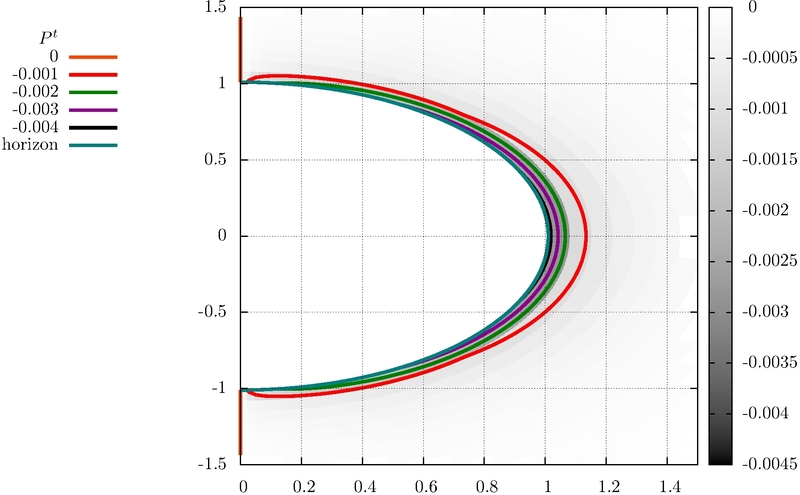}}
\caption{(color online) Behavior of Higgs and gauge fields in the extremal Kerr-Sen black hole 
background with charge $Q = 0.1 Q_{max}$ for the boson Higgs mass $m_{X} = 1.2$. }
\label{fig6}
\end{figure}
\begin{figure}[H]
\subfloat[$X$]{\includegraphics[scale=1.2]{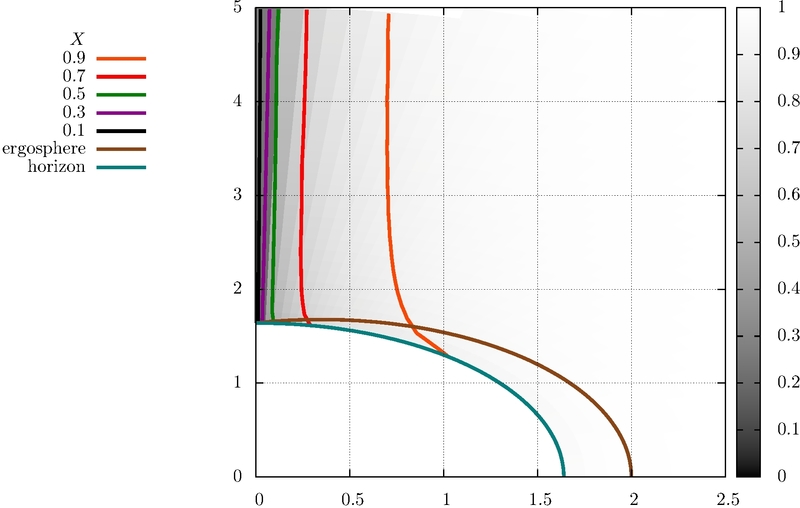}} \qquad
\subfloat[$P^{\varphi}$]{\includegraphics[scale=1.2]{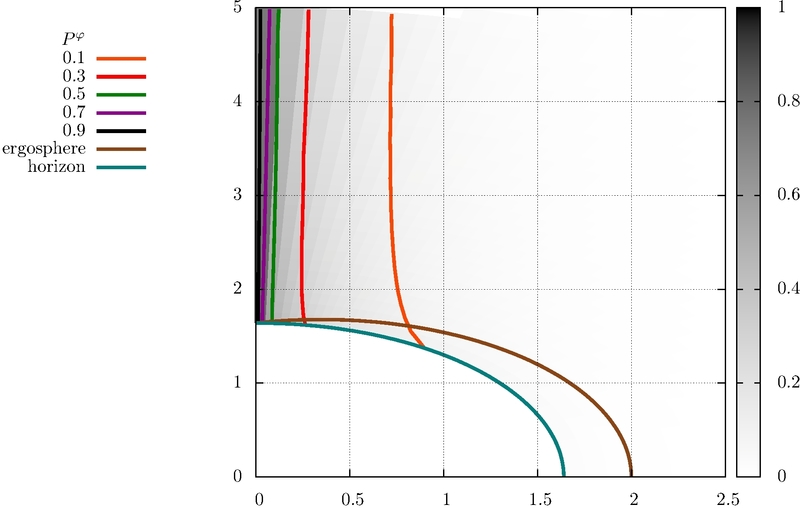}}
\\
\subfloat[$P^{t}$]{\includegraphics[scale=1.2]{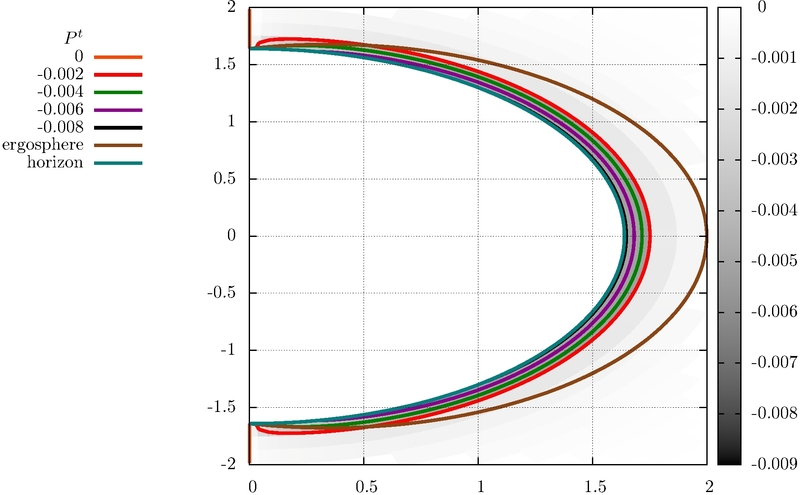}}
\caption{(color online) Behavior of Higgs and gauge fields in the spacetime of the extremal 
Kerr-Sen black hole with charge $Q = 0.8 Q_{max}$ for the boson Higgs mass $m_{X} = 4.0$. }
\label{fig7}
\end{figure}
\begin{figure}[H]
\subfloat[$X$]{\label{fig8a}\includegraphics[scale=1.2]{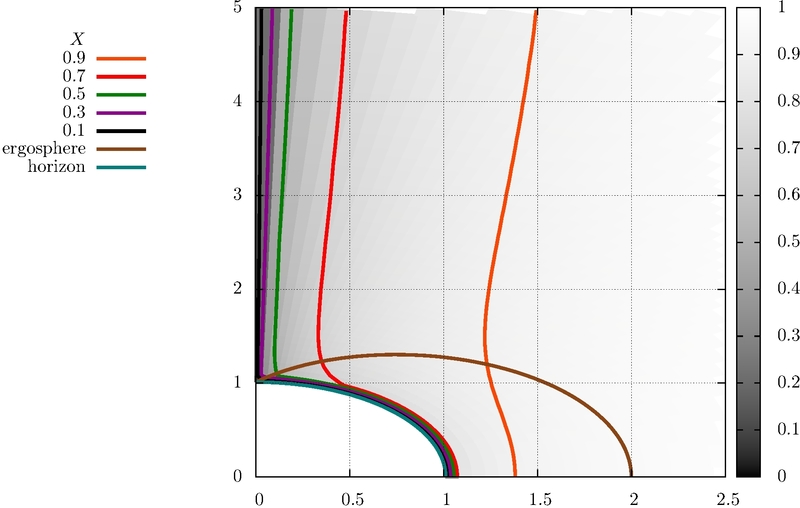}} \qquad
\subfloat[$P^{\varphi}$]{\label{fig8b}\includegraphics[scale=1.2]{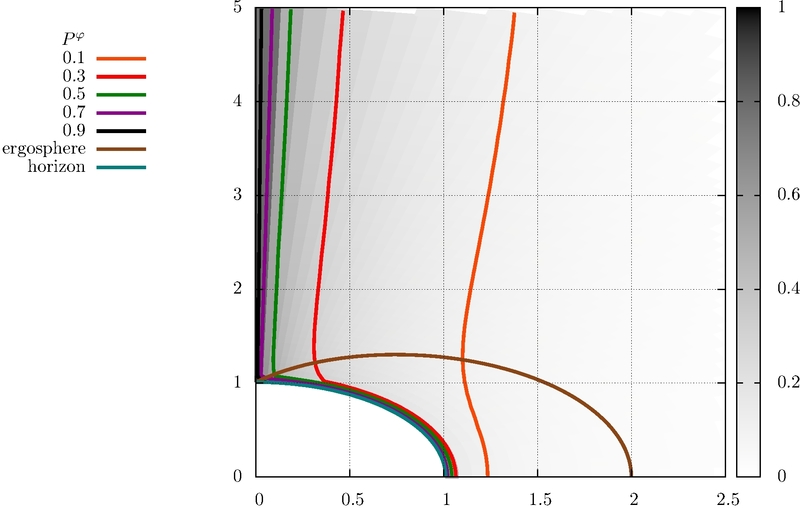}}
\\
\subfloat[$P^{t}$]{\label{fig8c}\includegraphics[scale=1.2]{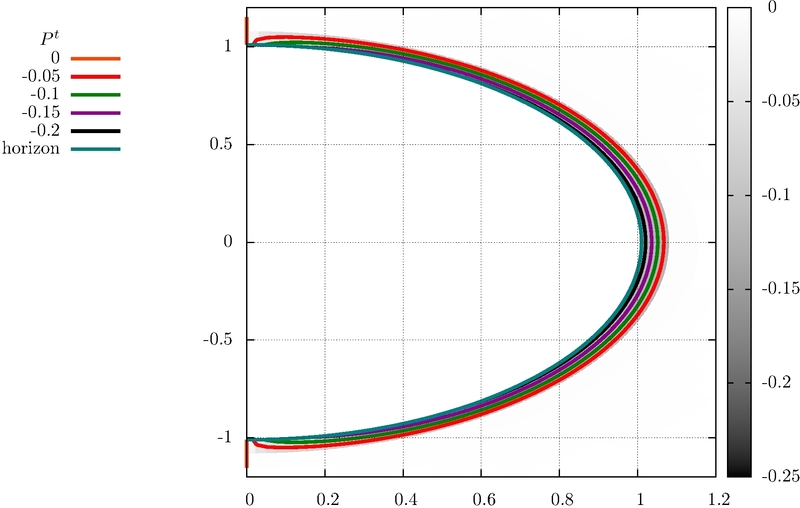}}
\caption{(color online) Behavior of the Abelian Higgs vortex fields in the background of the extremal Kerr-Sen black hole. 
We put black hole charge as $Q = 0.1 Q_{max}$, the boson Higgs mass $m_{X} = 0.6$ and the Bogomol'nyi parameter 
$\tbeta = 1$, respectively.}
\label{fig8}
\end{figure}
\begin{figure}[H]
\subfloat[]{\label{fig9a}\includegraphics[scale=1.2]{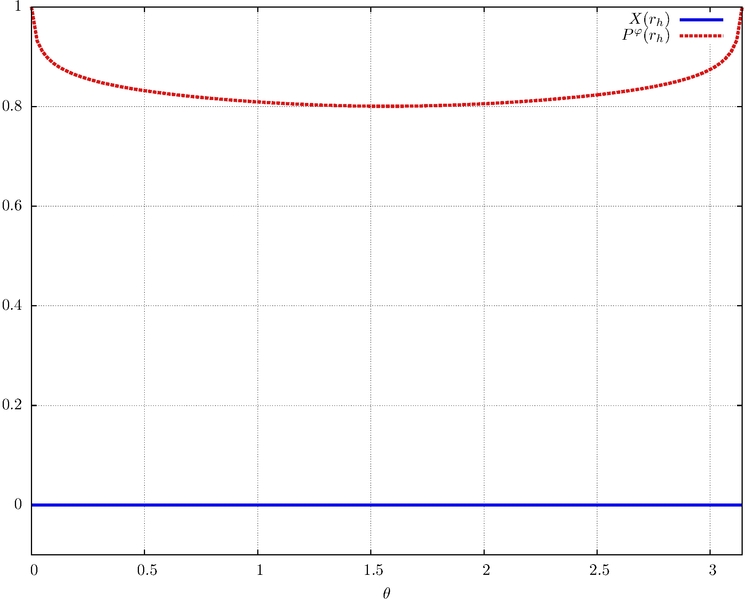}} \qquad
\subfloat[]{\label{fig9b}\includegraphics[scale=1.2]{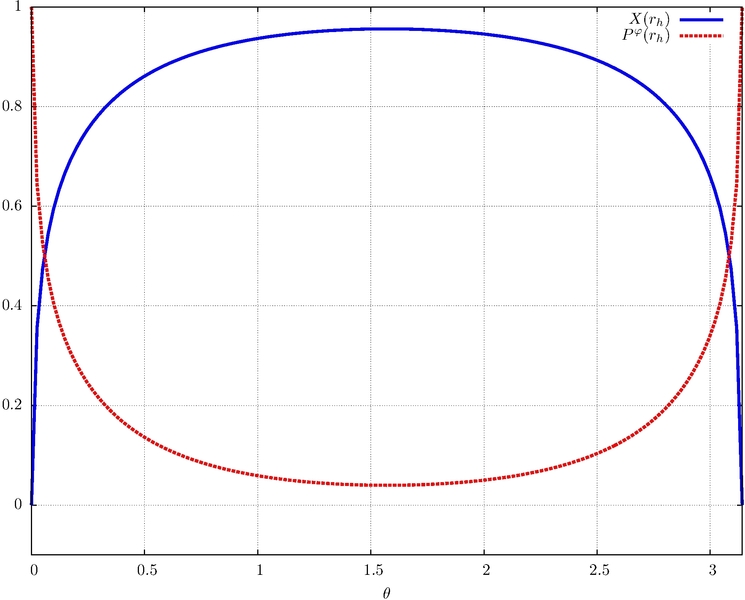}}
\caption{(color online) Profiles of the Higgs field $X$ and $P^{\varphi}$ gauge field component on
(a) the extremal Kerr-Sen black hole event horizon.
We set black hole charge  $Q = 0.1 Q_{max}$ and the boson Higgs mass $m_{X} = 0.6$.
(b) the nonextremal Kerr-Sen black hole horizon with the same charge, $a = 0.9 a_{max}$ and $m_{X} = 0.9$.}
\end{figure}

In the next step of our considerations we refine our studies to the case of the extremal Kerr-Sen
black hole vortex system, for which $a = a_{max}$.
In Fig.\ref{fig6} we plotted the behaviour of $X$ and $P^\alpha$ fields for Kerr-Sen black hole with
charge $Q = 0.1~Q_{max}$ and set Higgs boson mass as $m_{X} = 1.2$. Further, in
Fig.\ref{fig7} we studied the case with $Q = 0.8~Q_{max}$ and Higgs boson mass $m_{X} = 4.0$.
The comparisons of the obtained results reveal that the larger Higgs boson mass is the easier
vortex penetrates the black hole in question.\\
%%%%%%%%%%%%%%%%%%%%%%%%%
In Ref.\cite{gre13} it was claimed that in the case of Kerr vortex system, for small mass black 
hole one should have flux expulsion.
This situation corresponds to the case when we set $Q = 0$ in our code and choose $m_{X} < 1$. 
However, it turns out that for the Kerr-Sen black hole the problem is far more involved.
%%%%%%%%%%%%%%%%%%%%%%%%
Namely, in Fig.\ref{fig8} we displayed the results for {\it small} black hole with
$Q = 0.1~Q_{max}$,~$m_{X} = 0.6$ and $\tbeta = 1$. 
%%%%%%%%%%%%%%%%
In Fig.\ref{fig8a} it can be seen that Higgs field
$X$ is expelled from the black hole, while Fig.\ref{fig8c} indicate that $P^{t}$ field is concentrated in the small 
region around the black hole event horizon.
%%%%%%%%%%%%%%%
A closer look of the behaviour of the Higgs field $X$ and  
$P^{\varphi}$ gauge field on the horizon of
the black hole is presented in Fig.9a. It is instructive to compare the aforementioned
results with the reaction of $X$ and $P^{\varphi}$ fields in nonextremal Kerr-Sen background.
Therefore, in Fig.9b we depicted behaviour of the fields in question
for black hole with $a = 0.9$ and $m_{X} = 0.9$. 
%%%%%%%%%%%%%%%
Just on the basis of Fig.\ref{fig9a}, one has the situation when
the Higgs field is expelled from the event horizon, while the gauge flux penetrates the black hole interior.
%%%%%%%%%%%%%%
We obtain the case of terminating the Abelian Higgs vortex on the Kerr-Sen event horizon.
This situation was widely studied in the case of various static black hole Abelian Higgs vortex systems
\cite{ach95}-\cite{mod99}.
\par
\begin{figure}[H]
\subfloat[$X(Q = 0.1 Q_{max})$]{\label{fig10a}\includegraphics[scale=1.2]{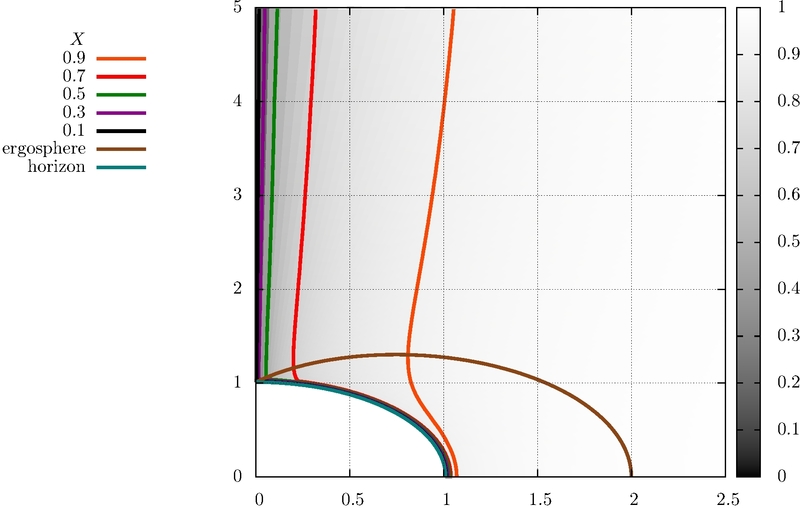}} \qquad
\subfloat[$P^{\varphi}(Q = 0.1 Q_{max})$]{\label{fig10b}\includegraphics[scale=1.2]{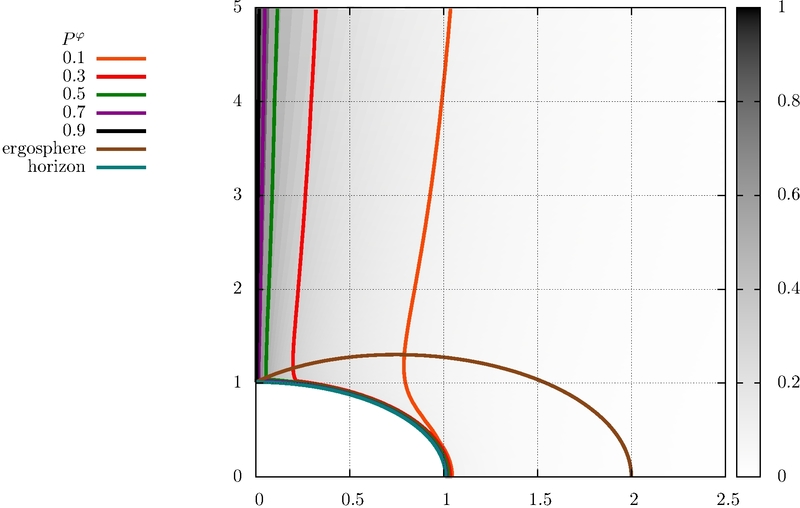}}
\\
\subfloat[$X(Q = 0.5 Q_{max})$]{\label{fig10c}\includegraphics[scale=1.2]{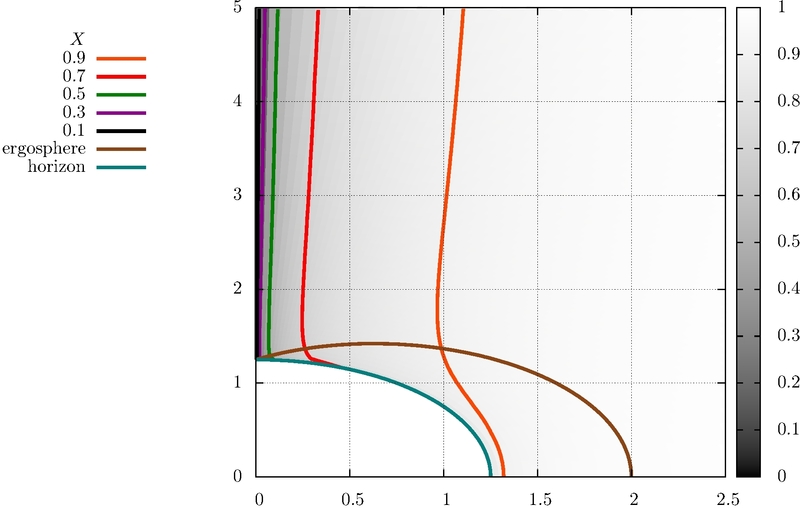}} \qquad
\subfloat[$P^{\varphi}(Q = 0.5 Q_{max})$]{\label{fig10d}\includegraphics[scale=1.2]{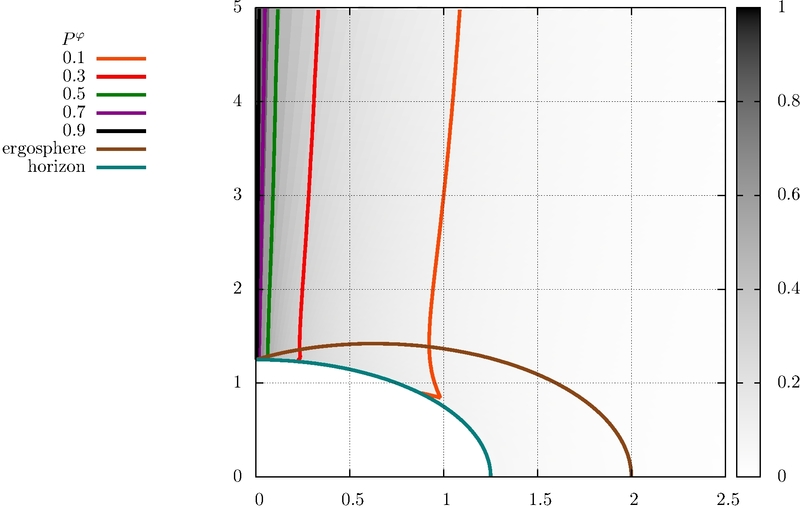}}
\\
\subfloat[$P^{t}(Q = 0.1 Q_{max})$]{\label{fig10e}\includegraphics[scale=1.2]{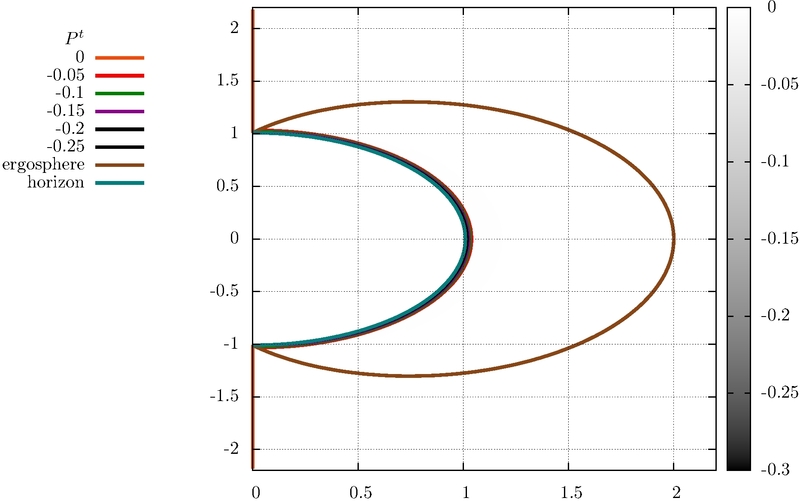}} \qquad
\subfloat[$P^{t}(Q = 0.5 Q_{max})$]{\label{fig10f}\includegraphics[scale=1.2]{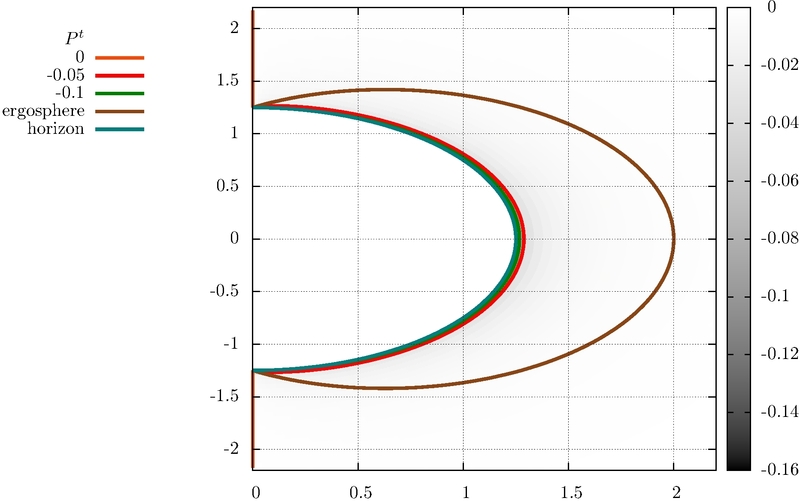}}
\caption{(color online) Behavior of Higgs and gauge fields in the spacetime
of the extremal Kerr-Sen black hole for two different charges. We set $Q = 0.1~Q_{max},~0.5~Q_{max}$ and
establish the boson Higgs mass as $m_{X} = 0.9$.}
\label{fig10}
\end{figure}

In Fig.\ref{fig10} we displayed pictures describing the behaviour of the Higgs field and gauge fields for
the extremal Kerr-Sen solution characterized by two different values of the charge, we set boson Higgs mass
as equal to $0.9$. Analyzing 
Fig.10a-Fig.10b and Fig.10c-Fig.10d 
we may conclude that the increase of the black hole charge 
changes the solutions from representing the expulsion to the solution depicting
piercing of the black hole by a vortex.
Consequently,
it can be remarked that the larger black hole charge is the harder is to obtain
the expulsion of Higgs and gauge fields from the extremal black hole. On the other hand,
in all studied cases $P^t$ component of gauge field is concentrated in the nearby of the black hole
event horizon. All these facts may may indicate that for $Q$ close to $Q_{max}$ there exists only the 
piercing vortex solution.\\
By virtue of the above, one can 
%%%%%%%%%%%%%%
state 
%%%%%%%%%%%%%%
that for the extremal Kerr-Sen black hole the
Abelian Higgs vortex can be expelled or can thread the black hole, depending on the 
values of $Q$ and $m_{X}$. Generally for the Higgs boson mass larger than
one the black hole is pierced by the vortex. The limit value of $m_{X} = 1$ roughly corresponds 
to the black hole with mass $M = 10^{8} kg$ 
%%%%%%%%%%%%%%%%%
for 
%%%%%%%%%%
Higgs particle mass of 126 GeV.  
In the special case of $Q = 0$ (Kerr black hole) for $m_{X} < 1$ the Higgs field and the flux 
of the gauge field are expelled from the black hole interior,
while for $m_{X}>1$ they penetrate the black event hole horizon \cite{gre13}.

%%%%%%%%%%%%%%%%%%%%%%%%%%%%%%%%%%%%%%%%%%%%%%%%%%%%%%%%%%%%%%%%%%%%%%%%%%%%%%%%%%%%%%%%%%%%%%%%%%%%
%%%%%%%%%%%%%%%%%%%%%%%%%%%%%%%%%%%%%%%%%%%%%%%%%%%%%%%%%%%%%%%%%%%%%%%%%%%%%%%%%%%%%%%%%%%%%%%%%%%%
\section{Conclusions}
In our paper we have resolved the problem concerning the existence of an Abelian Higgs
vortex on the stationary axisymmetric Kerr-Sen black hole background. We elaborate the problem in question
both analytically and numerically. Assuming the complete separation between degrees of freedom of
the Kerr-Sen black hole and an Abelian Higgs vortex, one justifies that the vortex can be regarded 
from infinity as hair on the black hole spacetime. Analytical calculations, where we use
the so-called {\it thin string limit}, i.e., mass of the considered black hole $M_{bh} \gg 1$, and taking into account
backreaction of the Abelian Higgs vortex on black hole geometry, reveal that vortex influence is far more subtle
than causing only a conical defect. The conical defect is felt not only by $\varphi$-coordinate but also by
$t$-coordinate. It leads to the new phenomenon of shifting the position of the ergosphere. Numerical calculations
disclose that the shifting is towards black hole event horizon and depends on black hole charge and
angular momentum parameter. Namely, the increase of the Kerr-Sen black hole charge
(decrease of the angular momentum parameter) induces the decline of the magnitude of the ergosphere displace.
\par
Numerical solutions of the equations of motion for Abelian Higgs vortex Kerr-Sen black hole system
proclaim that in the case of the nonextremal Kerr-Sen black hole it is always pierced by the Abelian Higgs vortex.
It turned out that the increase of the Higgs boson mass $m_X$
governed the decrease of the width of the vortex.
The angular momentum parameter of the black hole influences on the concentration of $P^t$ component of the gauge
field. Namely, the bigger value of the angular momentum parameter one considers the closer to the black hole 
event horizon the concentration in question takes place.
\par
On the other hand, in the case of the extremal Kerr-Sen black hole case, it happened that the bigger
Higgs boson mass one takes into account the easier the vortex penetrates the black hole
under consideration. For large boson Higgs masses
$m_X >1$ and for large value of black hole charge the extremal Kerr-Sen black hole in always threaded by the 
vortex. Contrary, for the small extremal black holes with
$Q \sim 0.1 Q_{max}$ and $m_{X} < 1$,
we obtain the solution when the Abelian Higgs vortex
ends on the black hole event horizon, i.e., the Higgs field is expelled while the gauge field penetrates
the black hole.\\
Moreover, the larger charge of the black hole taken into considerations the harder is to achieve the expulsion
of the vortex fields. As far as $P^t$ component of the gauge field is concerned it is concentrated
in the nearby of the Kerr-Sen black hole event horizon.

%%%%%%%%%%%%%%%%%%%%%%%%%%%%%%%%%%%%%%%%%%%%%%%%%%%%%%%%%%%%%%%%%%%%%%%%%%%%%%%%%%%%%%%%%%%%%%%%%%%%%%%
%%%%%%%%%%%%%%%%%%%%%%%%%%%%%%%%%%%%%%%%%%%%%%%%%%%%%%%%%%%%%%%%%%%%%%%%%%%%%%%%%%%%%%%%%%%%%%%%
%\begin{appendix}

%\section{Irred   } 
%\label{irtf}
%\end{appendix}
\begin{acknowledgments}
MR was partially supported by the grant of the National Science Center
$2011/01/B/ST2/00408$. LN was supported by the Polish National Center under doctoral scholarship
$2013/08/T/ST2/00122$.
\end{acknowledgments}
%%%%%%%%%%%%%%%%%%%%%%%%%%%%%%%%%%%%%%%%%%%%%%%%%%%%%%%%%%%%%%%%%%%%%%%%%%%%%%%%%%%%%%%%%%

%%%%%%%%%%%%%%%%%%%%%%%%%%%%%%%%%%%%%%%%%%%%%%%%%%%%%%%%%%%%%%%%%%%%%%%%%%%%%%%%%%%%%%%%%%%%%%%%%%%%%%%
%%%%%%%%%%%%%%%%%%%%%%%%%%%%%%%%%%%%%%%%%%%%%%%%%%%%%%%%%%%%%%%%%%%
%%%%%%%%%%%%%%%%%%%%%%%%%%%%%%%%%%%%%%%%%%%%%%%%%%%%%%%%%%%%%%%%%%%%%%%%%%%%%%%%%
%%%%%%%%%%%%%%%%%%%%%%%%%%%%%%%%%%%%%%%%%%%%%%%%%%%%%%%%%%%%%%%%%%%%%%%%%%%%%%%%%

%%%%%%%%%%%%%%%%%%%%%%%%%%%%%%%%%%%%%%%%%%%%%%%%%%%%%%%%%%%%%%%%%%%%%%%%%%%%%%%%%%%%%%%%%%%%%%%%%%%%%%%%%%%%%%

%%%%%%%%%%%%%%%%%%%%%%%%%%%%%%%%%%%%%%%%%%%%%%%%%%%%%%%%%%%%%%%%%%%%%%%%%%%%%%%%%%%%%%%%%%%%%%%%%%%%%%%%%%%%%
%%%%%%%%%%%%%%%%%%%%%%%%%%%%%%%%%%%%%%%%%%%%%%%%%%%%%%%%

%%%%%%%%%%%%%%%%%%%%%%%%%%%%%%%%%%%%%%%%%%%%%%%%%%%%%%%%%%%%%%%%%%%%%%%%%%%%%%%%%%%%%

%%%%%%%%%%%%%%%%%%%%%%%%%%%%%%%%%%%%%%%%%%%%%%%%%%%%%%%%%%%

%%%%%%%%%%%%%%%%%%%%%%%%%%%%%%%%%%%%%%%%%%%%%%%%%%%%%%%%%%%%%%%%%%%%%%%%%%%%%%%%%%%%%%%%%%%%%%%%%%%%%%%%%%%%%
%%%%%%%%%%%%%%%%%%%%%%%%%%%%%%%%%%%%%%%%%%%%%%%%%%%%%%%%%%%%%%%%%%%%%%%%%%%%%%%%%%%%%%%%%%%%%%%%%%

\begin{thebibliography}{99}
% 
\def\cmp#1#2#3{{ Commun. Math. Phys.} {\bf #1}, #2 (#3)}
\def\lmp#1#2#3{{ Lett. Math. Phys.} {\bf #1}, #2 (#3)}
\def\hpa#1#2#3{{ Hell. Phys. Acta} {\bf #1}, #2 (#3)}
\def\grg#1#2#3{{ Gen. Rel. Grav.} {\bf #1}, #2 (#3)}
\def\pr#1#2#3{{ Phys. Rev.} {\bf #1}, #2 (#3)}
\def\prl#1#2#3{{ Phys. Rev. Lett.} {\bf #1}, #2 (#3)}
\def\prc#1#2#3{{ Phys. Rev. C} {\bf #1}, #2 (#3)}
\def\prd#1#2#3{{ Phys. Rev. D} {\bf #1}, #2 (#3)}
\def\pl#1#2#3{{ Phys. Lett} {\bf #1}, #2 (#3)}
\def\pla#1#2#3{{ Phys. Lett. A} {\bf #1}, #2 (#3)}
\def\plb#1#2#3{{ Phys. Lett. B} {\bf #1}, #2 (#3)}
\def\prep#1#2#3{{ Phys. Reports} {\bf #1}, #2 (#3)}
\def\phys#1#2#3{{ Physica} {\bf #1}, #2 (#3)}
\def\jcp#1#2#3{{ J. Comput. Phys.} {\bf #1}, #2 (#3)}
\def\jmp#1#2#3{{ J. Math. Phys.} {\bf #1}, #2 (#3)}
\def\jpm#1#2#3{{ J. Phys. A: Math. Gen.} {\bf #1}, #2 (#3)}
\def\cpr#1#2#3{{ Computer Phys. Rept.} {\bf #1}, #2 (#3)}
\def\cqg#1#2#3{{ Class. Quantum Grav.} {\bf #1}, #2 (#3)}
\def\cma#1#2#3{{ Computers Math. Applic.} {\bf #1}, #2 (#3)}
\def\mc#1#2#3{{ Math. Compt.} {\bf #1}, #2 (#3)}
\def\apj#1#2#3{{ Astrophys. J.} {\bf #1}, #2 (#3)}
\def\apjs#1#2#3{{ Astrophys. J. Suppl.} {\bf #1}, #2 (#3)}
\def\acta#1#2#3{{ Acta Astronomica} {\bf #1}, #2 (#3)}
%%%%%%%%%%%%%%%%%%%%%%%%%%%%%%%%%%%%%%%%%%%%%%%%%%%%%%%%%%%%%%%%%%%%%%%%%%
\def\appb#1#2#3{{ Acta Phys. Polonica B} {\bf #1}, #2 (#3)}
\def\apl#1#2#3{{Ann. Physik. (Leipzig)} {\bf #1}, #2 (#3)}
\def\jetp#1#2#3{{JETP Lett.} {\bf #1}, #2 (#3)}

\def\anp#1#2#3{{Ann. Phys. } {\bf #1}, #2 (#3)}
\def\sa#1#2#3{{ Sov. Astro.} {\bf #1}, #2 (#3)}
\def\sia#1#2#3{{ SIAM J. Sci. Statist. Comput.} {\bf #1}, #2 (#3)}
\def\aa#1#2#3{{ Astron. Astrophys.} {\bf #1}, #2 (#3)}
\def\mnras#1#2#3{{ Mon. Not. R. astr. Soc.} {\bf #1}, #2 (#3)}
\def\npb#1#2#3{{ Nucl. Phys. B} {\bf #1}, #2 (#3)}
\def\prsla#1#2#3{{ Proc. R. Soc. London, Ser. A} {\bf #1}, #2 (#3)}
\def\jhep#1#2#3{{ JHEP} {\bf #1}, #2 (#3)}
\def\jgp#1#2#3{{ J.Geom.Phys.} {\bf #1}, #2 (#3)}
\def\livrev#1#2#3{{ Living Rev. Relativity} {\bf #1}, #2 (#3)}

\def\nuc#1#2#3{{Nuovo Cimento B } {\bf #1}, #2 (#3)}
\def\ijmp#1#2#3{{Int. J. Mod. Phys. D} {\bf #1}, #2 (#3)}
\def\ijmpe#1#2#3{{Int. J. Mod. Phys. E} {\bf #1}, #2 (#3)}
\def\atmp#1#2#3{{Adv. Theor. Math. Phys.} {\bf #1}, #2 (#3)}
\def\ptps#1#2#3{{Prog. Theor. Phys. Suppl.} {\bf #1}, #2 (#3)}
\def\ptp#1#2#3{{Prog. Theor. Phys. } {\bf #1}, #2 (#3)}
\def\lmp#1#2#3{{Lett. Math. Phys. } {\bf #1}, #2 (#3)}
\def\mmj#1#2#3{{Mich. Math. j. } {\bf #1}, #2 (#3)}
\def\rmpj#1#2#3{{Rev. Mod. Phys} {\bf #1}, #2 (#3)}
%
\def\hepph#1#2{{ hep-ph }{\bf #1} (#2)}
\def\hepth#1#2{{ hep-th }{\bf #1} (#2)}
\def\grqc#1#2{{ gr-qc }{\bf #1} (#2)}
\def\ibid#1#2#3{{ {\it ibid.} }{\bf #1}, #2 (#3)}
\def\asteri#1#2#3{{ Ast\'erisque }{\bf #1}, #2 (#3)}

%
%%%%%%%%%%%%%%%%%%%%%%%%%%%%%%%%%%%%%%%%%%%%%%%%%%%%%%%%%%%%%%%%%%%%%%
\bibitem{book}
M.Heusler, {\it Black Hole Uniqueness Theorems} 
(Cambridge: Cambridge University Press, 1996),\\
P.Chrusciel, J.L.Costa, and M.Heusler, \livrev{15}{7}{2012},\\
P.Chrusciel and J.L.Costa, \asteri{321}{195}{2008}.

\bibitem{rog12}
M.Rogatko, \prd{86}{064005}{2012}.

\bibitem{nd}
G.W.Gibbons, D.Ida, and T.Shiromizu, \prd{66}{044010}{2002},\\
G.W.Gibbons, D.Ida, and T.Shiromizu, \prl{89}{041101}{2002},\\
S.Hollands, A.Ishibashi, and R.M.Wald, \cmp{271}{699}{2007},\\
M.Rogatko, \cqg{19}{L151}{2002},\\
M.Rogatko, \prd{67}{084025}{2003},\\
M.Rogatko, \ibid{70}{044023}{2004},\\
M.Rogatko, \ibid{71}{024031}{2005},\\
M.Rogatko, \ibid{73}{124027}{2006}.

\bibitem{nrot}
Y.Morisawa and D.Ida, \prd{69}{124005}{2004},\\
Y.Morisawa, S.Tomizawa, and Y.Yasui, \ibid{77}{064019}{2008},\\
M.Rogatko, \ibid{70}{084025}{2004},\\
M.Rogatko, \ibid{77}{124037}{2008},\\
S.Hollands and S.Yazadjiev, \cmp{283}{749}{2008},\\
S.Hollands and S.Yazadjiev, \cqg{25}{095010}{2008},\\
D.Ida, A.Ishibashi, and T.Shiromizu, \ptps{189}{52}{2011},\\
S.Hollands and A.Ishibashi, \cqg{29}{163001}{2012}.
%%%%%%%%%%%%%%%%%%%%%%%%%%%%%%%%%%%%%%%%%%%%%%%%%%%%%%%%%%%%%%%%%%%%%%%%%%%%%%%%%%%
%%%%%%%%%%%%%%%%%%%%%%%%%%%%%%%%%%%%%%%%%%%%%%%%%%%%%%%%%%%%%%%%%%%%%%%%%%%%%%%%%%%%%%%%%%%%%%%%%%

\bibitem{sugra}
A.K.M.Massod-ul-Alam, \cqg{14}{2649}{1993},\\
M.Mars and W.Simon, \atmp{6}{279}{2003},\\
M.Rogatko, \cqg{14}{2425}{1997},\\
M.Rogatko, \prd{58}{044011}{1998},\\
M.Rogatko, \ibid{59}{104010}{1999},\\
M.Rogatko, \cqg{19}{875}{2002},\\
S.Tomizawa, Y.Yasui, and A.Ishibashi, \prd{79}{124023}{2009},\\
S.Tomizawa, Y.Yasui, and A.Ishibashi, \prd{81}{084037}{2010},\\
J.B.Gutowski, \jhep{0408}{049}{2004},\\
J.P.Gauntlett, J.B.Gutowski, C.M.Hull, S.Pakis, and H.S.Real, \cqg{20}{4587}{2003}.


%%%%%%%%%%%%%%%%%%%%%%%%%%%%%%%%%%%%%%%%%%%%%%%%%%%%%%%%%%%%%%%%%%%
\bibitem{adscft}
J.M.Maldacena, \atmp{2}{231}{1998},\\
S.A.Hartnoll, \cqg{26}{224002}{2009},\\ 
G.T.Horowitz, \cqg{28}{114008}{2011},\\
G.T.Horowitz, {\it Introduction to Holographic Superconductors}, \hepth{1002.1722}{2010}.


\bibitem{shi12}
T.Shiromizu, S.Ohashi, and R.Suzuki, \prd{86}{064041}{2012}.

\bibitem{bak13}
B.Bakon and M.Rogatko, \prd{87}{084065}{2013}.

\bibitem{shi13}
T.Shiromizu and S.Ohashi, \prd{87}{087501}{2013}.



%%%%%%%%%%%%%%%%%%%%%%%%%%%%%%%%%%%%%%%%%%%%%%%%%%%%%%%%%%%%%%%%%%%%%%%%%%%%%%%%%%%
\bibitem{shi13a}
T.Shiromizu and K.Tanabe, \prd{87}{081504}{2013}.

\bibitem{rog13}
M.Rogatko, \prd{88}{024051}{2013}.


\bibitem{ba88}
R.Bartnik and J.McKinnon, \prl{61}{41}{1988}.

\bibitem{biz90}
P.Bizon, Phys.Rev.Lett.{\bf 64}, 2844 (1990).
\bibitem{kun90}
H.P.K\"unzle and A.K.M.Masood-ul-Alam, \jmp{31}{928}{1990}.
\bibitem{str90}
N.Straumann and Z.H.Zhou, \plb{237}{353}{1990},\\
N.Straumann and Z.H.Zhou, \ibid{243}{33}{1990}.
\bibitem{biz91}
P.Bizon and R.M.Wald, \plb{267}{173}{1991}.


\bibitem{vil}
A.Vilenkin and E.P.S.Shallard, {\it Cosmic Strings and Other Topological Defects},
Cambridge University Press, Cambridge (1994).
\bibitem{ary86}
M.Aryal, L.H.Ford, and A.Vilenkin, \prd{34}{2263}{1986}. 
\bibitem{dow92}
F.Dowker, R.Gregory, and J.Traschen, \prd{45}{2762}{1992}.
\bibitem{mod98}
R.Moderski and M.Rogatko, \prd{57}{3449}{1998}.
\bibitem{ach95}
A.Achucarro, R.Gregory, and K.Kuijken, \prd{52}{5729}{1995}.
\bibitem{cha}
A.Chamblin, J.Ashbourn-Chamblin, R.Emparan, and A.Sornborger, \prd{58}{124014}{1998},\\
A.Chamblin, J.Ashbourn-Chamblin, R.Emparan, and A.Sornborger, \prl{80}{4378}{1998},\\
F.Bonjour, R.Emparan, and R.Gregory, \prd{59}{084022}{1999}.
\bibitem{mod99}
R.Moderski and M.Rogatko, \prd{58}{124016}{1998},\\
R.Moderski and M.Rogatko, \ibid{60}{104040}{1999}.

\bibitem{gre92}
R.Gregory and J.A.Harvey, \prd{46}{3302}{1992}.

\bibitem{nak11}
L.Nakonieczny and M.Rogatko, \prd{84}{044029}{2011}.
\bibitem{nak12}
L.Nakonieczny and M.Rogatko, \prd{85}{044013}{2012}.

\bibitem{ghe02}
A.Ghezelbash and R.Mann, \prd{65}{124022}{2002}.
\bibitem{gre13}
R.Gregory, D.Kubiznak, and D.Wills, {\it Rotating black hole hair}, \grqc{1303.0519}{2013}.

\bibitem{rog10}
M.Rogatko, \prd{82}{044017}{2010}.

\bibitem{gal95}
D.Gal'tsov, A.Garcia, and O.Kechkin, \jmp{36}{5023}{1995}.
\bibitem{gar92}
D.Garfinkle, G.T.Horowitz, and A.Strominger, \prd{43}{3140}{1991},\\
D.Garfinkle, G.T.Horowitz, and A.Strominger, \ibid{45}{3888}{1992}.
\bibitem{sen92}
A.Sen, \prl{69}{1006}{1992}.

\bibitem{nil73}
H.B.Nielsen and P.Olesen, \npb{61}{45}{1973}.

\bibitem{gib13}
G.W.Gibbons, A.H.Mujtaba, C.N.Pope, \cqg{30}{125008}{2013}.


\bibitem{Press}
W. H. Press, S. A. Teukolsky, W. T. Vetterling, and B. P. Flannery,
{\it Numerical Recipes in C},
Cambridge University Press, Cambridge  (1992). 


\end{thebibliography}
\end{document}